\documentclass[submission,copyright,creativecommons]{eptcs}
\providecommand{\event}{PLACES 2022} % Name of the event you are submitting to
\usepackage{breakurl}             % Not needed if you use pdflatex only.
\usepackage{underscore}           % Only needed if you use pdflatex.
\usepackage{xparse}
\usepackage{booktabs}   %% For formal tables:
\usepackage{subcaption} %% For complex figures with subfigures/subcaptions
\usepackage[dvipsnames]{xcolor}

\usepackage{amsmath,amsfonts,amsthm} % http://texdoc.net/texmf-dist/doc/latex/amsmath/amsldoc.pdf. ,amssymb
\usepackage[capitalize]{cleveref}
\usepackage{footnote}
\makesavenoteenv{tabular}
\usepackage{thm-restate}
\usepackage{stmaryrd} % \fatsemi among others
\usepackage{cmll}
\usepackage{xspace}
\usepackage{multicol}
\usepackage[inline]{enumitem}
\usepackage{rotating}
\usepackage{prooftree} % use mathpartir
\usepackage{mathpartir}
\usepackage{tikz}
\usetikzlibrary{automata, positioning, arrows}
\tikzset{
        ->,  % makes the edges directed
        >=stealth, % makes the arrow heads bold
        node distance=4.5em, % specifies the minimum distance between two nodes. C
        initial text=$ $, % sets the text that appears on the start arrow
        }

\newcommand{\typec}[1]{{\color{RoyalBlue}{#1}}} % c for colour
\newcommand{\termc}[1]{{\color{RedOrange}{#1}}}
\newcommand{\kindc}[1]{\termc{#1}} % kind
\newcommand{\tsymc}[1]{{\color{OliveGreen}{#1}}} % terminal symbol
\newcommand{\nsymc}[1]{{\color{BrickRed}{#1}}} % nonterminal symbol
\newcommand{\grmeq}{\; ::= \;}
\newcommand{\grmor}{\; \mid \;}

\newcommand{\keyword}[1]{\ensuremath{\mathsf{#1}}\xspace}
\newcommand{\tkeyword}[1]{\keyword{\typec{#1}}}

\newcommand{\symkeyword}[1]{\keyword{\tsymc{#1}}}

\newcommand{\Label}[1]{\tkeyword{#1}}
\newcommand{\nill}{\Label{Nil}}
\newcommand{\consl}{\Label{Cons}}
\newcommand{\nodel}{\Label{Node}}
\newcommand{\leafl}{\Label{Leaf}}
\newcommand{\intlistt}{\tkeyword{IntList}}
\newcommand{\inttreet}{\tkeyword{IntTree}}

\newcommand{\intreet}{\tkeyword{InputTree}}
\newcommand{\htreet}{\tkeyword{HTree}}
\newcommand{\procname}[1]{\mathtt{#1}}
\newcommand{\sendint}{\procname{sendInt}}

\newcommand{\unitl}{\symkeyword{unit}}
\newcommand{\ldomain}{\tsymc{\shortrightarrow_d}}
\newcommand{\lrange}{\tsymc{\shortrightarrow_r}}

\newcommand{\operator}[1]{\operatorname{#1}}
\newcommand{\subexpr}{\operator{subterms}}

\newcommand{\Eq}{\doteq} % The equal symbol in equations
\newcommand{\Empty}{\varepsilon} % The empty sequence (context, for example)
\newcommand{\teq}{\simeq} % type equivalence
\newcommand{\gequiv}{\approx} % grammar bisimulation

\newcommand{\Oplus}{\hspace{-.1ex}\oplus\hspace{-.1ex}} % internal choice with a better spacing

\newcommand{\incr}[1]{{#1}^{+1}}
\newcommand{\N}{\mathcal{N}}

\newcommand{\pcal}{\mathcal{P}}
\newcommand{\R}{\mathcal{R}}

\newcommand{\T}{\mathcal{T}}
\newcommand{\foralllabel}[2]{(\forall{#1}\in{#2})}
\newcommand{\foralllinL}{\foralllabel \ell L}

\newcommand{\declrel}[2]{#1\hfill\fbox{{#2}}}

\newcommand{\inductive}{(\emph{inductive})\xspace}
\newcommand{\coinductive}{(\emph{coinductive})\xspace}

\newcommand{\skind}{\kindc{\textsc s}}
\newcommand{\tkind}{\kindc{\textsc t}}
\newcommand{\vkind}{\kindc\kappa}
\newcommand{\vkindp}{\vkind'}

\newcommand{\End}{\tkeyword{end}}
\newcommand{\Skip}{\tkeyword{skip}}
\newcommand{\INp}{?}
\newcommand{\INn}[1]{\typec{\INp{#1}}}		% no trailing dot
\newcommand{\OUTp}{!}
\newcommand{\OUTn}[1]{\typec{\OUTp{#1}}}	% no trailing dot

\newcommand{\variantf}[3]{\langle{#1}\colon {#2}\rangle_{{#1}\in{#3}}} % Full
\newcommand{\variant}[3]{\variantf{#1}{#2_#1}{#3}}
\newcommand{\variants}[1]{\langle{#1}\rangle}
\newcommand{\recordf}[3]{\{{#1}\colon {#2}\}_{{#1}\in{#3}}} % Full
\newcommand{\records}[1]{\{#1\}}
\newcommand{\record}[3]{\recordf{#1}{{#2}_{#1}}{#3}} % Records indexed by sets L of labels, used for internal/external choice types and case processes
\newcommand{\varrecf}[3]{\llparenthesis{#1}\colon {#2}\rrparenthesis_{{#1}\in{#3}}}
\newcommand{\varrecs}[1]{\llparenthesis{#1}\rrparenthesis}
\newcommand{\varrec}[3]{\varrecf{#1}{{#2}_{#1}}{#3}}
\newcommand{\intchoicep}{\Oplus}
\newcommand{\intchoice}{\typec\intchoicep} % internal choice
\newcommand{\extchoicep}{\&}
\newcommand{\extchoice}{\typec\extchoicep}	% external choice
\newcommand{\choicep}{\odot}
\newcommand{\choice}{\typec\choicep}
\newcommand{\semit}[2]{\typec{{#1};{#2}}}
\newcommand{\dott}[2]{\typec{{#1}.{#2}}} %dot version of semi

\newcommand{\unit}{\tkeyword{unit}}
\newcommand{\dualoft}{\tkeyword{dualof}}
\newcommand{\Int}{\tkeyword{int}}

\newcommand{\function}[2]{\typec{#1 \rightarrow #2}}

\newcommand{\quant}{\exists\!\forall}

\newcommand{\forallt}[2]{\typec{\forall^{\kindc{#1}}{#2}}}

\newcommand{\TT}{\typec{T}}
\newcommand{\UT}{\typec{U}}

\newcommand{\Endl}{\keyword{end}}

\newcommand{\judgementlabel}[1]{\mathrm{#1}} % labels in judgements

\newcommand{\judgement}[2]{{#1} \: \judgementlabel{#2}}

\newcommand{\judgementrel}[3]{{#1} \; {#2} \; {#3}}

\newcommand{\judgementrelctxlab}[4]{{#1} \vdash \judgementrel{#2}{#3}{#4}}

\newcommand{\istype}[3]{\judgementrelctxlab{#1}{\typec{#2}}{:}{\kindc{#3}}}
\newcommand{\isdone}[1]{\judgement{\typec{#1}}{\!\checkmark}}
\newcommand{\isnotdone}[1]{\judgement{\typec{#1}}{\not\!\checkmark}}

\newcommand{\isEquiv}[3]{\judgementrel{\typec{#1}}{#2}{\typec{#3}}}
\DeclareDocumentCommand{\isequiv} { m m } { \isEquiv{#1}{\teq}{#2} }
\newcommand{\isequivnokind}[2]{\isEquiv{#1}{\teq}{#2}}
\newcommand{\isnotequiv}[2]{\isEquiv{#1}{\not\teq}{#2}}

\DeclareDocumentCommand{\isbisim} { m m } { \isEquiv{#1}{\sim}{#2} }

\newcommand{\iseqt}[2]{\judgementrel{\typec{#1}}{\Eq}{\typec{#2}}} % type equation
\newcommand{\entry}[2]{\typec{#1}\colon\kindc{#2}}

\newcommand{\ltsarrow}[1]{\stackrel{#1}{\longrightarrow}}
\newcommand{\gltsred}[3]{\nsymc{#1}\ltsarrow{\tsymc{#2}}\nsymc{#3}}
\newcommand{\ltsred}[3]{\typec{#1}\ltsarrow{\tsymc{#2}}\typec{#3}}

\newcommand{\iscontrt}[1]{\judgement{\typec{#1}}{contr}} % types
\newcommand{\prodsymbol}{\rightarrow}

\newcommand{\gprod}[2]{\nsymc{#1} \prodsymbol #2}

\ifx\vv\undefined
\newcommand{\vv}[1]{\marginpar{\textcolor{blue}{#1}}}
\else
\renewcommand{\vv}[1]{\marginpar{\textcolor{blue}{#1}}}
\fi

\newcommand{\myparagraph}[1]{\vspace{-1em}\paragraph{#1}}
\newcommand{\ie}{i.e.,\xspace} % comma
\newcommand{\etal}{et al.\xspace} % al abbreviates aliae, no need for am \emph,
\newcommand{\skiprulename}{Skip}

\newcommand{\seqrulename}{Seq}

\newcommand{\idrulename}{Id}

\newcommand{\unitrulename}{Unit}
\newcommand{\arrowrulename}{Arrow}

\newcommand{\indexrulename}{Index}

\newcommand{\rulename}[1]{\text{\small\sc #1}\xspace}

\newcommand{\kindrulename}[1]{\rulename{K-{#1}}}
\newcommand{\kunit}{\kindrulename{\unitrulename}}

\newcommand{\kindex}{\kindrulename{\indexrulename}}

\newcommand{\kskip}{\kindrulename{\skiprulename}}

\newcommand{\kid}{\kindrulename{\idrulename}}
\newcommand{\typeequivrulename}[1]{\rulename{E-{#1}}}

\newcommand{\eqvarl}{\typeequivrulename{\idrulename L}}
\newcommand{\eqvarr}{\typeequivrulename{\idrulename R}}

\newcommand{\eqskip}{\typeequivrulename{\skiprulename}}

\newcommand{\eqskipseql}{\typeequivrulename{SkipSeqL}}
\newcommand{\eqskipseqr}{\typeequivrulename{SkipSeqR}}

\newcommand{\eqvarseql}{\typeequivrulename{IdSeqL}}
\newcommand{\eqvarseqr}{\typeequivrulename{IdSeqR}}
\newcommand{\eqseqseql}{\typeequivrulename{SeqSeqL}}
\newcommand{\eqseqseqr}{\typeequivrulename{SeqSeqR}}

\newcommand{\equnit}{\typeequivrulename{\unitrulename}}
\newcommand{\eqfun}{\typeequivrulename{\arrowrulename}}

\newcommand{\eqindex}{\typeequivrulename{\indexrulename}}
\newcommand{\eqindexseqonel}{\typeequivrulename{{\indexrulename}Seq1L}}
\newcommand{\eqindexseqoner}{\typeequivrulename{{\indexrulename}Seq1R}}
\newcommand{\eqindexseqtwo}{\typeequivrulename{{\indexrulename}Seq2}}

\newcommand{\termrulename}[1]{\rulename{$\checkmark$-{#1}}}
\newcommand{\teskip}{\termrulename{\skiprulename}}

\newcommand{\teseq}{\termrulename{\seqrulename}}
\newcommand{\teid}{\termrulename{\idrulename}}
\newcommand{\premspace}{\quad\;}
\title{Higher-order Context-free Session Types in System F}

\author{Diana Costa\quad Andreia Mordido\quad Diogo Poças\quad Vasco T. Vasconcelos
\institute{LASIGE, Faculdade de Ciências, Universidade de Lisboa, Portugal}
\email{dfdcosta,afmordido,dmpocas,vmvasconcelos@ciencias.ulisboa.pt}
}
\def\titlerunning{Higher-order Context-free Session Types in System F}
\def\authorrunning{D. Costa, A. Mordido, D. Poças \& V.T. Vasconcelos}

\begin{document}
\maketitle
% !TeX root = main.tex

\begin{abstract}
  We present an extension of System F with higher-order context-free session
  types. The mixture of functional types with session types has proven to be a
  challenge for type equivalence formalization: whereas functional type
  equivalence is often rule-based, session type equivalence usually follows a
  semantic approach based on bisimulations. We propose a unifying approach that
  handles the equivalence of functional and session types together. We present
  three notions of type equivalence: a syntactic rule-based version, a semantic
  bisimulation-based version, and an algorithmic version by reduction to the
  problem of bisimulation of simple grammars. We prove that the three notions
  coincide and derive a decidability result for the type equivalence problem of
  higher-order context-free session types.

% We provide a unifying approach for semantic, syntactic and
% algorithmic versions of type equivalence. propose 
% an algorithm to decide whether two types are equivalent. The 
% algorithm is based on a translation of higher-order session types
% into simples grammars. We provide results of decidability, termination,
% soundness and completeness.
\end{abstract}

%%% Local Variables:
%%% mode: latex
%%% TeX-master: "main"
%%% End:
% !TeX root = main.tex

\section{Introduction}

Session types describe the behaviour of structured
communication~\cite{DBLP:conf/concur/Honda93,DBLP:conf/esop/HondaVK98,DBLP:conf/parle/TakeuchiHK94}.
The behaviour of process $\sendint$ can be expressed by the session type
$\function{\function{\Int}{\dott{\OUTn \Int}{b}}}{b}$ asserting that the process
is given a value of type $\Int$ and a channel of type $\dott{\OUTn \Int}{b}$ and
returns a channel of type $\typec{b}$. Session types also provide primitives for
offering and selecting choices as well as for unbounded behaviour via recursion.
A client willing to send a list of integers on a channel can be governed by type
$\iseqt {\intlistt} {\intchoice\records{\consl\colon \dott{\OUTn
      \Int}{\intlistt}, \nill\colon \Endl }}$ stating that the client can either
choose $\consl$ or $\nill$. In the former case the client must subsequently send
an integer value and go back to the choice; in the latter case the protocol is
terminated as identified by type $\End$.

Traditional session types have proven particularly useful in the specification
of protocols of different natures, provided that they can be characterized by
regular languages. Traditional session types are restricted to tail
recursion---this specificity is not just a feature, it is rather a limitation:
there are numerous protocols whose traces cannot be characterized by regular
languages.
Context-free session types liberate session types from tail recursion. In the
context-free world, clients can send integer trees on channels in a type-safe
way, without requiring the exchange of additional channels:
$\iseqt {\inttreet} {\intchoice\records{\nodel\colon
    \semit{\inttreet}{\semit{\OUTn \Int}{\inttreet}}, \leafl\colon \Skip}}$.
Context-free session types provide a sequential composition operator $\typec{;}$
and the corresponding neutral element, $\Skip$. Governed by type $\inttreet$,
the client is now able to select $\nodel$, send the left subtree, followed by an
integer value, followed by the right subtree, as witnessed by the double
recursion on type identifier $\inttreet$.
The increase in the expressivity of types comes at a price: checking type
equivalence becomes a challenge. Thiemann and Vasconcelos proved that type
equivalence is decidable for context-free session
types~\cite{DBLP:conf/icfp/ThiemannV16}, but a practical type equivalence
algorithm was only provided a few years later by Almeida
\etal\cite{DBLP:conf/tacas/AlmeidaMV20}.

% For example, checking whether type $\semit{\inttreet}{\INn\Int}$ is
% equivalent to type
% $\intchoice\records{\nodel\colon
%     \semit{\inttreet}{\semit{\OUTn \Int}{\semit\inttreet{\INn\Int}}},
%     \leafl\colon \INn\Int}$ boils down to checking whether the grammars
% starting from nonterminal symbols $\nsymc{X}$ and $\nsymc{Y}$ are bisimilar:
% \begin{align*}
%   \nsymc{X} &\prodsymbol \nsymc{X_1}\,\nsymc{X_2}
%   &
%   \nsymc{X_1} &\prodsymbol \tsymc{\Oplus\keyword{Node}}\, \nsymc{X_1}\,\nsymc{X_3}\,\nsymc{X_1}
%   &
%   \nsymc{X_1} &\prodsymbol \tsymc{\Oplus\keyword{Leaf}}
%   &
%   \nsymc{X_3} &\prodsymbol \tsymc{!\keyword{int}}
%   & 
%   \nsymc{X_2} &\prodsymbol \tsymc{?\keyword{int}}
%   \\
%   \nsymc{Y} &\prodsymbol \tsymc{\Oplus\keyword{Node}}\,\nsymc{X_1}\,
%   \nsymc{X_3}\,\nsymc{X_1}\,
%   \nsymc{X_2}
%   &
%   \nsymc{Y} &\prodsymbol \tsymc{\Oplus\keyword{Leaf}}\,\nsymc{X_2}
% \end{align*}

All proposals in the literature consider first-order context-free session types:
decidability of type equivalence is only guaranteed when basic types (or any
other types that can be syntactically compared for equality) are exchanged in
messages.
This paper promotes context-free session types to the higher-order setting. In
this new setting we can define trees with values of non-basic types, such as the
type of a channel on which a binary tree of input-$\Int$ channels can be sent:
$$\iseqt {\intreet} {\intchoice\records{\nodel\colon
    \semit{\intreet}{\semit{\OUTn{(\INn\Int)}}{\intreet}}, \leafl\colon
    \Skip}}.$$

Higher-order context-free session types can be endowed with impredicative
polymorphism.
However, some care must be exercised. Allowing polymorphism over arbitrary types
may raise complications: should one consider
$\typec{\forall\alpha.(\alpha;\OUTn\Int)}$ a bona fide type? It really depends
on what we replace $\typec{\alpha}$ with: if $\Skip$, then we get a genuine type
$\typec{\Skip;\OUTn\Int}$; if $\unit$, then we get a bogus type
$\typec{\unit;\OUTn\Int}$. In order to distinguish functional types from session
types in the presence of polymorphic types we introduce kinds: $\tkind$ for
functional and $\skind$ for session types, collectively know as $\vkind$. The
universal type is then annotated with the kind of the bound variable,
$\typec{\forall\alpha\colon\vkind.T}$.

Nominal bound variables cause problems when checking type
equivalence. As such, we elide them and use De Bruijn
indices~\cite{debruijn:1972:lambda} to refer to polymorphic variables: the above
type is now written $\forallt{\skind}{0;\OUTn\Int}$, where $\typec{0}$ denotes
a type variable bound by the first enclosing $\typec{\forall}$.
Regardless of the nature of the bound variable, should the polymorphic type
itself be a session or a functional type? The answer to this question dictates
how one composes the type to form larger types. Currently we allow functional
polymorphism only.
% in a named syntax, type $\forallt{\tkind}{\function{\INn0}{0}}$ would be
% written as $\typec{\forall\alpha\colon\tkind.\function{\INn\alpha}{\alpha}}$.

In order to check the equivalence of polymorphic context-free types we reduce
the problem of checking type equivalence into that of checking the bisimilarity
of simple grammars, along the lines of Almeida
\etal\cite{DBLP:conf/tacas/AlmeidaMV20}. We have implemented the procedure for
checking type equivalence for the monomorphic fragment of the language of this
paper in a branch of the FreeST compiler~\cite{freest}. For example, the simple grammar
associated with type $\intreet$ is as follows.
\begin{align*}
	   \nsymc{X} &\prodsymbol \tsymc{\Oplus\keyword{Node}}\,\nsymc{X}\,\nsymc{X_1}\,\nsymc{X}
	   &
	   \nsymc{X} &\prodsymbol \tsymc{\Oplus\keyword{Leaf}}
	   &
	   \nsymc{X_1} &\prodsymbol \tsymc{\OUTp_d}\,\nsymc{X_2}\nsymc{\bot}
	   &
	   \nsymc{X_1} &\prodsymbol \tsymc{\OUTp_c}
       \\
       \nsymc{X_2} &\prodsymbol \tsymc{\INp_{\!d}}\,\nsymc{X_{3}}\nsymc{\bot}
       &
       \nsymc{X_2} &\prodsymbol \tsymc{\INp_c}
       &
       \nsymc{X_3} &\prodsymbol \tsymc{\keyword{int}}
\end{align*}

This work explores the type equivalence problem for higher-order context-free
session types. The main contributions are:
\begin{itemize}
\item A new formulation of context-free session types that allows a clean
  integration of session types and functional types, so that session types may
  exchange both session types and conventional functional types (such as
  functions or records);
\item A syntactic and a semantic definition of equivalence for higher-order
  session types---a rule-based approach and a labelled transition system---which
  we prove to coincide;
%\item a proof that type equivalence for higher-order context-free
%  session types is decidable;
\item A type equivalence algorithm by reduction to the bisimilarity of simple
  grammars, as well as results of termination, soundness, completeness and
  decidability of type equivalence.
\end{itemize}

%%% Local Variables:
%%% mode: latex
%%% TeX-master: "main"
%%% End:
% !TeX root = main.tex

\section{Polymorphic higher-order context-free session types}
%\section{System F with higher-order context-free session types}
\label{sec:types}

In this section we introduce an extension of System
F~\cite{girard:1971:extension,DBLP:conf/programm/Reynolds74} with higher-order
context-free session types. We rely on non-negative numerals---denoted
$\typec m$ and $\typec n$---to describe polymorphic variables; a set
$\mathbb{L}$ of labels---denoted by $\typec k$ and $\typec \ell$---to specify
labelled choices, records and variant types; and type identifiers---denoted
$\typec X$ and $\typec Y$---to provide for recursive types. A kinding system
distinguishes session types (denoted by kind $\skind$) from functional types
(denoted by $\tkind$). We use symbol~$\vkind$ to denote either $\skind$ or
$\tkind$. Kinds for bound variables are kept in a kinding context $\Delta$
containing bindings of the form $\entry{n}{\vkind}$. The type formation rules
are presented in \cref{fig:type-formation}.

% !TeX root = main.tex
\begin{figure}[t!]
Polarity, view, records and quantifiers\qquad
	\begin{align*}
      \typec\sharp \grmeq{} \typec? \grmor{} \typec!
      &&
      \choice \grmeq{} \extchoice \grmor \intchoice
      &&
      \typec{\varrecs{\cdot}} \grmeq{} \typec{\records{\cdot}} \grmor \typec{\variants{\cdot}}
      &&
      \typec\quant \grmeq{} \typec\forall \grmor \typec\exists
    \end{align*}
  \begin{multicols}{2}
  \declrel{Is-terminated \inductive}{$\isdone T$}
  \begin{mathpar}
    \infer[\teskip]{\isdone\Skip}{}
    \and 
    \infer[\teseq]{
      \isdone{T}
      \premspace
      \isdone{U}
    }{
      \isdone{\semit TU}
    }
    \and 
    \infer[\teid]{
      \iseqt XT
      \premspace
      \isdone T
    }{
      \isdone{X}
    }
  \end{mathpar}
  \qquad Kind and kinding environment\hfill{}
  \begin{align*}
    \vkind &\grmeq \skind \grmor \tkind
    \\
    \Delta &\grmeq \Empty \grmor \Delta, \typec n \colon \vkind
    \\
    \Delta^{+1} &\;\,=\;\; \{\typec{n+1} \colon \vkind \mid \typec n \colon \vkind\in \Delta\}
    % \vkind \grmeq{}& \pkind \grmor \karrow {\pkind}K \grmor \karrow K\pkind
  \end{align*}
\end{multicols}
\declrel{Contractivity \inductive}{$\iscontrt T$}
  \begin{mathpar}
    \infer[\rulename{C-Axiom}]{\parbox{4.5cm}{$\typec T=\unit,\function{T}{U},\typec{\varrec{\ell}{T}{L}}\\\Skip,\typec{\sharp T},\typec{\choice\record \ell T L},\typec n$}}{\iscontrt{T}}
    \quad
    \infer[\rulename{C-Quant}]{\iscontrt T}{\iscontrt{\quant^\vkind T}}{}
%    \and 
%    \infer[\rulename{C-Exists}]{\iscontrt T}{\iscontrt{\exists^\vkind T}}{}
    \quad
    \infer[\rulename{C-Seq1}]{
      \isdone T
      \premspace
      \iscontrt U
    }{
      \iscontrt{T;U}
    }
    \quad 
    \infer[\rulename{C-Seq2}]{
      \isnotdone T
      \premspace
      \iscontrt T
    }{
      \iscontrt{T;U}
    }
    \quad 
    \infer[\rulename{C-Id}]{
      \iseqt XT
      \premspace
      \iscontrt T
    }{
      \iscontrt X
    }
  \end{mathpar}
  \declrel{Type formation \coinductive}{$\istype {\Delta} T \vkind$}
  \begin{mathpar}
    % FUNCTIONAL TYPE CONSTRUCTORS
    \infer[\kunit]{\istype {\Delta} \unit \tkind}{}
    \quad\;\;
    \infer[\rulename{K-Arrow}]{
      \istype {\Delta} {T} \vkind 
      \quad 
      \istype {\Delta} {V} {\vkind'}
    }{
      \istype {\Delta} {\function TV} \tkind
    }
    \quad\;\;
    \infer[\rulename{K-Rcd}]{
      \istype {\Delta} {T_\ell} {\vkind_\ell}
      \quad
      \foralllinL %\text{ for all }\ell\in L
    }{
      \istype {\Delta} {\varrecf \ell T L} \tkind
    }
    \quad\;\; 
    \infer[\rulename{K-Quant}]{
      \istype {\incr\Delta, \typec 0\colon \vkind} {T} {\vkind'}
    }{
      \istype {\Delta} {\quant^\vkind T} {\tkind}
    }
    % \quad\;\;
    % \infer[\rulename{K-Variant}]{
    % \istype {\Delta} {T_\ell} \tkind \text{ for all }\ell\in L
    % }{
    %   \istype {\Delta} {\record \ell T L} \tkind
    % }
    \quad\;\;
    % SESSION TYPE CONSTRUCTORS
    \infer[\kskip]{\istype {\Delta} \Skip \skind}{}
    \\
    \infer[\rulename{K-Msg}]{\istype {\Delta} T \vkind}{\istype {\Delta} {\typec\sharp T} \skind}
    \quad
    \infer[\rulename{K-Choice}]{
      \istype {\Delta} {T_\ell} \skind
      \quad
      \foralllinL %\text{ for all }\ell\in L
    }{
      \istype \Delta {\choice \record \ell T L} \skind
    }
    \quad
    \infer[\rulename{K-Seq}]{
      \istype {\Delta} T \skind
      \premspace
      \istype {\Delta} U \skind
    }{
      \istype {\Delta} {\semit{T}{U}} \skind
    }
    \quad
    % POLYMORPHIC TYPE CONSTRUCTORS
    \infer[\kindex]{
      \typec n\colon\vkind\in{\Delta}
    }{
      \istype {\Delta} n \vkind
    }
    % \quad 
    % \infer[\rulename{K-ExtChoice}]{
    % \istype {\Delta} {T_\ell} \skind \text{ for all }\ell\in L
    % }{
    %   \istype \Delta {\extchoice \record \ell T L} \skind
    % }
    %   \quad 
    %   \infer[\rulename{K-Exists}]{
    %   \istype {\incr\Delta, \typec 0\colon \vkind} {T} {\vkind'}
    % }{
    %   \istype {\Delta} {\exists^\vkind T} {\tkind}
    % }
    \quad
    \infer[\kid]{
      \iseqt XT
      \premspace
      \iscontrt T
      \premspace
      \istype {\Delta} T \vkind
    }{
      \istype {\Delta} X \vkind
    } 
  \end{mathpar}
  \caption{Type formation}
  \label{fig:type-formation}
\end{figure}

%%% Local Variables:
%%% mode: latex
%%% TeX-master: "main"
%%% End:

The first four rules in the figure introduce functional types: the
$\typec{\unit}$ type, functions $\typec{\function TV}$, records
$\typec{\record \ell T L}$, variants $\typec{\variant \ell T L}$ and
polymorphic types $\typec{\forall^\vkind T}$. This restriction on polymorphic
types is crucial to ensure that the translation in \cref{sec:grammars} is well
defined and indeed maps types to simple grammars.
De Bruijn indices (starting at $\typec 0$) allow checking polymorphic types
against polymorphic types without needing to worry about the concrete names of
variables.
% To provide for a 
% To ease notation and avoid the constant invocation of $\alpha$-renaming, we
% remove the identification of the bound variables alongside the quantifiers.
% Instead, we consider De Bruijn indices (starting at $\typec 0$), respecting the
% order of enclosing quantifiers (as suggested, for instance, by
% Pierce~\cite{DBLP:books/daglib/0005958} for $\lambda$-terms).
To ensure the correct formation of types, quantifiers are annotated with the
kind of the bound variable---denoted by the superscript $\vkind$. With this
notation, instead of writing
$\typec{\forall \alpha\colon\tkind.\,\function{\alpha}{\forall \beta\colon
    \skind.\, \function{\semit{\OUTn{\alpha}}{\beta}}{\beta}}}$ for the
$\keyword{send}$ primitive, we write
$\typec{\forall^\tkind\function{0}{\forall^\skind\function{\semit{\OUTn{1}}{0}}{0}}}$.
When crossing a quantifier, the numerals in the kinding context
$\Delta$ are incremented by 1---denoted by the superscript $+1$ in rule
\rulename{K-Quant}.

The next four rules in \cref{fig:type-formation} introduce session types: the
$\Skip$ type, output of arbitrary types $\OUTn{T}$, input of arbitrary types
$\INn{T}$, internal choice $\typec {\intchoice \record \ell T L}$, external
choice $\typec {\extchoice \record \ell T L}$ and the sequential composition
$\semit{T}{U}$.
The last two rules in the figure introduce numerals $\typec n$ as polymorphic type
variables and type identifiers $\typec X$, defined by equations of the form
$\iseqt XT$.

A signature $\Sigma$ is a finite collection of equations $\iseqt XT$ where no
type identifier $\typec X$ occurs twice at the left of an equation. 
Whenever a signature $\Sigma$ is clear from context, we write
$\typec X\doteq \typec T$ to mean an entry in $\Sigma$.
The right-hand sides of equations must be \emph{contractive}. Contractivity ensures
that a type eventually rewrites to a type constructor, eschewing non-types such
as $\typec X$ with equations $\iseqt XY$ and $\iseqt YX$. A more elaborate
example of a (non-contractive, hence) non-type is $\typec X$ with
$\iseqt X{\semit \Skip Y}$ and $\iseqt Y{\semit XY}$.

The notion of contractivity relies on the \emph{is-terminated} predicate, which
is true on types comprising only $\Skip$, sequential composition and type
identifiers.
Terminated types have a simple characterisation, which justifies the inclusion of predicate
$\isnotdone T$ in rule \rulename{C-Seq2}~\cite{DBLP:conf/tacas/AlmeidaMV20}.
The is-terminated and contractivity predicates are inductively defined,
whereas type formation is coinductive.

We can easily show that kinds are unique: any syntactic object has at most one
kind, in which case we call the object a type.
We say that $\typec T$ is a type when there are $\Delta$ and $\vkind$ such that $\istype{\Delta}{T}{\vkind}$.

%%% Local Variables:
%%% mode: latex
%%% TeX-master: "main"
%%% End:
% !TeX root = main.tex

\section{Syntactic type equivalence}
\label{sec:syntacticequivalence}

The rules for type equivalence are shown in \cref{fig:type-equivalence}. The
novelty lies in the rules for sequential composition. Intuitively, sequential
composition has a monoidal
structure---$\isequivnokind{\semit{(\semit TU)}V}{\semit T{(\semit UV)}}$---with
$\Skip$ being the (left and right)
neutral element---$\isequivnokind{\semit\Skip T}{\semit T \Skip}\isequivnokind{}T$. In
addition, sequential composition must distribute with
choice---$\isequivnokind{\semit{\intchoice\record\ell T
    L}U}{\intchoice\recordf{\ell}{\semit{T_\ell}{U}}{L}}$.
The first eight rules, from \equnit to \eqindex, are the congruence rules for all
type constructors (\ie without type identifiers and sequential composition). Rules \eqvarl
and \eqvarr interpret recursive types equi-recursively.
The rules in the last three lines are the rules for sequential composition. For
each session type constructor $\typec T$ one finds a left rule (of the form
$\isequiv{\semit{T}{U}}{V}$) and a right rule $(\isequiv{V}{\semit{T}{U}}$).
Since sequential composition does not distribute with message passing or indices,
we require an additional rule for these constructors (rules \rulename{E-MsgSeq2} 
and \rulename{E-IndexSeq2}).
As we are using a coinductive proof scheme, we have rules that
% that entail these properties by
`move' the sequential composition operator `down' the syntax (or, to put it in
another way, that `move' type constructors that actually produce something `up'
the syntax). This is why for types of the form $\semit TU$ we look at the
structure of $\typec T$ to decide which rule to apply next.

% !TeX root = main.tex
\begin{figure}[t!]
  \declrel{Type equivalence \coinductive}{$\isequiv TT$}
  \begin{mathpar}
    % CONGRUENCE (functional)
    \infer[\equnit]{\isequiv\unit\unit}{}
    \quad
    \infer[\eqfun]{
      \isequiv TU
      \premspace
      \isequiv VW
    }{
      \isequiv{\function TV}{\function UW}
    }
    \quad
    \infer[\rulename{E-Rcd}]{
      \isequiv{T_\ell}{U_\ell}
      \\
      \foralllinL % \text{ for all }\ell\in L
    }{
      \isequiv{\varrecf \ell {T_\ell} L}{\varrecf \ell {U_\ell} L}
    }
    \quad
    \infer[\rulename{E-Quant}]{
      \isequiv TU
    }{
      \isequiv {\quant^{\vkind} T}{\quant^{\vkind} U}
    }
    % \\
    \quad
    % CONGRUENCE (sessions)
    \infer[\eqskip]{
      \isequiv\Skip\Skip
    }{}
    \quad
    \infer[\rulename{E-Msg}]{
      \isequiv TU
    }{
      \isequiv {\sharp T}{\sharp U}
    }
    \\
    \infer[\rulename{E-Choice}]{
      \isequiv{T_\ell}{U_\ell}
      \qquad
      \foralllinL % \text{ for all }\ell\in L
    }{
      \isequiv{\choice\recordf\ell {T_\ell} L}{\choice\recordf\ell {U_\ell} L}
    }
    \quad
    % CONGRUENCE (polymorphic)
    % 
    \infer[\eqindex]{\isequiv nn}{}
    \quad
    \infer[\eqvarl]{
      \iseqt{X}{T}
      \premspace \iscontrt{T}
      \premspace \isequiv{T}{U}
    }{
      \isequiv{X}{U}
    }
    \quad
    \infer[\eqvarr]{
      \iseqt{X}{U}
      \premspace \iscontrt{U}
      \premspace \isequiv{T}{U}
    }{
      \isequiv{T}{X}
    }
    \\
    % Congruence (Semicolon)
    % Skip
    \infer[\eqskipseql]{
      \isequiv TU
    }{
      \isequiv {\semit \Skip T} U
    }
    \quad
    \infer[\eqskipseqr]{
      \isequiv TU
    }{
      \isequiv T {\semit\Skip U}
    }
    \quad
    % Message
    \infer[\rulename{E-MsgSeq1L}]{
      \isequiv TU \premspace \isdone V
    }{
      \isequiv{\semit {\sharp T} V}{\sharp U}
    }
    \quad
    \infer[\rulename{E-MsgSeq1R}]{
      \isequiv TU \premspace \isdone V
    }{
      \isequiv{\sharp T}{\semit {\sharp U} V}
    }
    \quad
    \infer[\rulename{E-MsgSeq2}]{
      \isequiv{T}{U} \premspace \isequiv{V}{W}
    }{
      \isequiv{\semit{\sharp T}{V}}{\semit{\sharp U}{W}}
    }
    \quad
    % Choice
    \infer[\rulename{E-ChoiceSeqL}]{
      \isequiv{\choice\recordf \ell {\semit{T_\ell}{U}} L}{V}
    }{
      \isequiv {\semit{\choice\recordf{\ell}{T_\ell}{L}}U}{V}
    }
    \\
    \infer[\rulename{E-ChoiceSeqR}]{
      \isequiv{U}{\choice\recordf \ell {\semit{T_\ell}{V}} L}
    }{
      \isequiv {U}{\semit{\choice\recordf{\ell}{T_\ell}{L}}V}
    }
    \quad
    % Semicolon
    \infer[\eqseqseql]{
      \isequiv {\semit T {(\semit UV)}}{W}
    }{
      \isequiv {\semit {(\semit TU)} V}{W}
    }
    \quad
    \infer[\eqseqseqr]{
      \isequiv {T}{\semit U {(\semit VW)}}
    }{
      \isequiv {T}{\semit {(\semit UV)} W}
    }
    \quad
    \infer[\eqindexseqonel]{
      \isdone T
    }{
      \isequiv{\semit nT}{n}
    }
    \quad
    \infer[\eqindexseqoner]{
      \isdone T
    }{
      \isequiv{n}{\semit nT}
    }
    \\
    \infer[\eqindexseqtwo]{
      \isequiv TU
    }{
      \isequiv{\semit nT}{\semit nU}
    }
    \quad
    \infer[\eqvarseql]{
      \iseqt{X}{T}
      \premspace \iscontrt{T}
      \premspace \isequiv{\semit TV}{U}
    }{
      \isequiv {\semit XV}{U}
    }
    \quad
    \infer[\eqvarseqr]{
      \iseqt{X}{U}
      \premspace \iscontrt{U}
      \premspace \isequiv{T}{\semit UV}
    }{
      \isequiv {T}{\semit XV}
    }
  \end{mathpar}
\caption{Type equivalence}
\label{fig:type-equivalence}
\end{figure}

\begin{restatable}[Agreement for type equivalence]{theorem}{typeagreement}
\label{thm:typeagreement}
  If 
  $\istype \Delta T\vkind$, $\istype\Delta U{\vkindp}$ and
  $\isequiv TU$, then $\vkind=\vkindp$.
\end{restatable}

Type equivalence is unkinded (yet we call it type equivalence). There are objects
in the equivalence relation that are not types. Object $\typec{\Skip;\unit}$ is
equivalent to $\unit$ yet it is not a type, that is, there are no $\Delta$ and
$\vkind$ such that $\istype{\Delta}{\Skip;\unit}{\vkind}$. This precludes a
stronger agreement result, namely, $\isequiv TU$ implies $\istype \Delta T\vkind$.

\begin{restatable}[Equivalence relation]{theorem}{equivalencerelation}
\label{thm:teq-equivalence}
$\teq$ is an equivalence relation on types.
\end{restatable}

\section{Semantic type equivalence}
\label{sec:semanticequivalence}

Following Gay and Hole~\cite{DBLP:journals/acta/GayH05}, we build on a type
bisimulation to provide a semantic definition for type equivalence. For this
purpose, we extend the original labelled transition system for context-free
session types~\cite{DBLP:conf/tacas/AlmeidaMV20,DBLP:conf/icfp/ThiemannV16} and
introduce labelled transitions for functional and higher-order types. The
definition of the labelled transition system (LTS) is in \cref{fig:lts}.

We start with functional types. Type $\unit$ transitions to $\Skip$ via label
$\unitl$. Function types induce two transitions: one via label
$\ldomain$ to the domain of the function, the other via
$\lrange$ to the range. Records and variants step to each
component $\typec k$ via labels $\tsymc{\records{}_k}$ and
$\tsymc{\variants{}_k}$, respectively. Polymorphic types transition via the
respective label, $\tsymc\forall^\vkind$ or $\tsymc\exists^\vkind$, to its body.

For session types, choices follow the original
proposal~\cite{DBLP:conf/icfp/ThiemannV16} and step via $\tsymc{\intchoicep_k}$
and $\tsymc{\extchoicep_k}$ to the continuation type, for each labelled choice
$\typec k$. However, message exchanges for higher-order types now feature two
distinct transitions: one to the type exchanged in the message (via label
$\tsymc{\OUTp_d}$, $\tsymc d$ for data) and the other to the continuation type
(via label $\tsymc{\OUTp_c}$, $\tsymc c$ for continuation).
Type $\Skip$ does not exhibit any transition.
Indices $\typec n$ transition by label $\tsymc n$ to type
$\Skip$ (rule \rulename{L-Index}) and type identifiers inherit the transitions
from the associated type (rule \rulename{L-Id}).
Finally, sequential composition $\semit{T}{U}$ distinguishes cases for all type
constructors in $\typec T$ (rules \rulename{L-SkipSeq} to \rulename{L-IdSeq}).

% !TeX root = main.tex

\begin{figure}[t!]
  Transition labels
  \begin{equation*}
	\tsymc a \grmeq
    \unitl \grmor
    \ldomain \grmor \lrange \grmor
    \tsymc{\varrecs{\cdot}_\ell} \grmor
    \tsymc\quant^\vkind \grmor
	\tsymc{\sharp_d} \grmor \tsymc{\sharp_c} \grmor
    \tsymc{\odot_\ell} \grmor
    % \tsymc{\choice_\ell} \grmor % yields a blue \odot
    \tsymc{n}
  \end{equation*}
  \declrel{Labelled transition system \inductive}{$\ltsred{T}{a}{U}$}
  \begin{mathpar}
    % FUNCTIONAL
  \infer[\rulename{L-Unit}]{
    \ltsred{\unit}{\unitl}{\Skip}
  }{}
  \quad\;\;
  \infer[\rulename{L-Arrow1}]{
    \ltsred{\function{T}{U}}{\ldomain}{T}
  }{}
  \quad\;\;
  \infer[\rulename{L-Arrow2}]{
    \ltsred{\function{T}{U}}{\lrange}{U}
  }{}
  \quad\;\;
  \infer[\rulename{L-Rcd}]{
    k\in L
  }{
    \ltsred{\varrecf{\ell}{T}{L}}{\varrecs{\,}_k}{T_k}
  }
  \quad\;\;
  \infer[\rulename{L-Quant}]{
    \ltsred{\quant^{\vkind}T}{\quant^{\vkind}}{T}
  }{}
  \quad\;\;
  % SESSION TYPES
  \infer[\rulename{L-Msg1}]{
    \ltsred{\sharp T}{\sharp_d}{T}
  }{}
  \\
  \infer[\rulename{L-Msg2}]{
    \ltsred{\sharp T}{\sharp_c}{\Skip}
  }{}
  \quad
  \infer[\rulename{L-Choice}]{
    k\in L
  }{
    \ltsred{\choice\recordf{\ell}{T_\ell}{L}}{\choicep_k}{T_k}
  }
  \quad
  \infer[\rulename{L-Index}]{
    \ltsred{n}{n}{\Skip}
  }{}
  \quad
  \infer[\rulename{L-Id}]{
    \iseqt{X}{T}
    \premspace
    \ltsred{T}{a}{U}
  }{
    \ltsred{X}{a}{U}
  }
  \quad
  % SEMICOLON
  \infer[\rulename{L-SkipSeq}]{
    \ltsred{T}{a}{U}
  }{
    \ltsred{\semit{\Skip}{T}}{a}{U}
  }
  \quad
  \infer[\rulename{L-MsgSeq1}]{
    \ltsred{\semit{\sharp T}{U}}{\sharp_d}{T}
  }{}
  \\
  \infer[\rulename{L-MsgSeq2}]{
    \ltsred{\semit{\sharp T}{U}}{\sharp_c}{U}
  }{}
  \quad
  \infer[\rulename{L-ChoiceSeq}]{
    k\in L
  }{
    \ltsred{\semit{\choice\recordf{\ell}{T_\ell}{L}}{U}}{\choicep_k}{\semit{T_k}{U}}
  }
  \quad
  \infer[\rulename{L-SeqSeq}]{
    \ltsred{\semit{T}{(\semit UV)}}{a}{W}
  }{
    \ltsred{\semit{(\semit TU)}{V}}{a}{W}
  }
  \quad
  \infer[\rulename{L-IndexSeq}]{
    \ltsred{\semit{n}{U}}{n}{U}
  }{}
  \quad
  \infer[\rulename{L-IdSeq}]{
    \iseqt{X}{T}
    \premspace
    \ltsred{\semit TU}{a}{V}
  }{
    \ltsred{\semit XU}{a}{V}
  }
  \end{mathpar}
  \caption{Labelled transition system}
  \label{fig:lts}
\end{figure}

%%% Local Variables:
%%% mode: latex
%%% TeX-master: "main"
%%% End:

%  \quad
%  \infer[\rulename{L-In1}]{
%    \ltsred{\INn T}{\INp_d}{T}
%  }{}
%  \quad
%  \infer[\rulename{L-In2}]{
%    \ltsred{\INn T}{\INp_c}{\Skip}
%  }{}
%  \qquad
%  \infer[\rulename{L-Variant}]{
%    k\in L
%  }{
%    \ltsred{\variant{\ell}{T}{L}}{\variants{}_k}{T_k}
%  }
%  \qquad
%  \infer[\rulename{L-ExtChoice}]{
%    k\in L
%  }{
%    \ltsred{\extchoice\recordf{\ell}{T_\ell}{L}}{\extchoicep_k}{T_k}
%  }
  % \\
  % \axiom{
  %   \ltsred{\tabs KT}{\tabs{K}{}}{T}
  % }
%  \and
%  \infer[\rulename{L-Exists}]{
%    \ltsred{\exists^{\vkind}T}{\exists^{\vkind}}{T}
%  }{}
%  \quad
%  \infer[\rulename{L-InSeq1}]{
%    \ltsred{\semit{\INn T}{U}}{\INp_d}{T}
%  }{}
%  \quad
%  \infer[\rulename{L-InSeq2}]{
%    \ltsred{\semit{\INn T}{U}}{\INp_c}{U}
%  }{}
%  \and
%  \infer[\rulename{L-ExtChoiceSeq}]{
%    k\in L
%  }{
%    \ltsred{\semit{\extchoice\recordf{\ell}{T_\ell}{L}}{U}}{\extchoicep_k}{\semit{T_k}{U}}
%  }
  % \\
  % \axiom{
  %   \ltsred{\semit{\forall^\vkind T}{U}}{\forall^{\vkind}_1}{T}
  % }
  % \\
  % \axiom{
  %   \ltsred{\semit{\forall^\vkind T}{U}}{\forall^{\vkind}_2}{U}
  % }
  % \\
  % \axiom{
  %   \ltsred{\semit{\exists^\vkind T}{U}}{\exists^{\vkind}_1}{T}
  % }
  % \\
  % \axiom{
  %   \ltsred{\semit{\exists^\vkind T}{U}}{\exists^{\vkind}_2}{U}
  % }

The labelled transition system does not preserve kinding. There are types in
transition relation whose (only) kinds do not match. One example is $\unit$ of
kind $\tkind$ that transitions to $\Skip$ of kind $\skind$.

A bisimulation is defined in the usual way from the labelled transition
system~\cite{sangiorgi2014introduction}.
We say that a type relation $\mathcal R$ is a \emph{bisimulation} if for
all $(\typec{T}, \typec{U}) \in \mathcal R$ and for
all~$\tsymc{a}$ we have:
\begin{enumerate}
	\item for each $\typec{T'}$ with $\ltsred T a{T'}$, there is $\typec{U'}$
	  such that $\ltsred Ua{U'}$ and $(\typec{T'}, \typec{U'}) \in \mathcal R$, and
	\item for each $\typec{U'}$ with $\ltsred U a {U'}$, there is $\typec{T'}$ such that $\ltsred T a{T'}$ and $(\typec{T'}, \typec{U'}) \in \mathcal R$.
\end{enumerate}
We say that two types are bisimilar, $\typec{T}\sim \typec{U}$, if there
is a bisimulation~$\mathcal R$ such that $(\typec{T}, \typec{U}) \in \mathcal R$.
%Alternatively, $\sim$ is the largest bisimulation.

We can easily check that the type for a function $\keyword{send}$ that first
receives the type of the message, then sends the value and only afterwards
receives the type for the continuation,
$\typec{\forall^\tkind\function{0}{\forall^\skind\function{\semit{\OUTn{1}}{0}}{0}}}$,
is not equivalent to the type of a function $\keyword{send'}$ that starts by
receiving the type of value to be exchanged and the type of the continuation
channel,
$\typec{\forallt{\tkind}{\forallt{\skind}{\function{1}{\function{\semit{\OUTn{1}}{0}}{0}}}}}$.
We have that
$
\typec{\forall^\tkind\function{0}{\forall^\skind\function{\semit{\OUTn{1}}{0}}{0}}}
\not\sim
\typec{\forallt{\tkind}{\forallt{\skind}{\function{1}{\function{\semit{\OUTn{1}}{0}}{0}}}}}
$ because, even if both types exhibit a transition by label
$\tsymc\forall^\tkind$, only the first type then exhibits a transition by
label~$\tsymc0$.

We can easily check that type bisimulation is deterministic (hence finitely
branching) and image finite. It features infinite transition sequences, as well
as transition sequences that visit infinitely many different
states~\cite{DBLP:conf/icfp/ThiemannV16}.

\begin{restatable}[Soundness and completeness]{theorem}{syntaxsemantic}
\label{thm:syntaxsemantic} Let $\TT$, $\UT$ be types. Then $\isequiv TU$ iff $\isbisim TU$.
    %iff $\lcal(\nsymc{X}_{\typec{T}})=\lcal(\nsymc{X}_{\typec{U}})$.
\end{restatable}

The proviso that $\TT$ and $\UT$ are types is important. Objects
$\typec{\unit;\Skip}$ and $\Skip$ are bisimilar (no transition applies to
either), yet they cannot be shown equivalent.

\section{Bisimulation for simple grammars}
\label{sec:grammars}

% In this section, we provide a type equivalence algorithm 
% by reduction to \emph{simple grammars}. We provide results 
% on termination, soundness and completeness. We also provide
% a complexity analysis for several fragments of our type system.

% \paragraph{Simple grammars.}
A grammar is given by a tuple $(\T, \N, \nsymc{X}, \R)$ where: $\T$ is a set of
terminal symbols, denoted by $\tsymc{a},\tsymc{b},\tsymc{c}$, $\N$ is a set of
nonterminal symbols, denoted by $\nsymc{X}, \nsymc{Y}, \nsymc{Z}$, nonterminal
$\nsymc{X}\in \N$ is the starting symbol and
$\pcal \subseteq \N \times (\T \cup \N)^\ast$ is a set of productions.
Greek letters $\sigma$ and $\tau$ denote (possibly empty) words of terminal and
nonterminal symbols;
greek letters $\nsymc\gamma$ and $\nsymc\delta$ denote (possibly empty) words of
nonterminal symbols only.
Each production is written as $\gprod{X}{\sigma}$.
% where $\nsymc{X}$ is a
% nonterminal and $\sigma$ is a word of terminals and nonterminals.
%
It is well-known that every grammar can be converted into an equivalent grammar
in Greibach normal form~\cite{greibach:1965:normalform}. A grammar is in
Greibach normal form if $\pcal\subseteq \N \times \T \times \N^\ast$, in other
words, when every production is of the form $\gprod{X}{\tsymc{a}\nsymc{\gamma}}$.
% where $\tsymc{a}$ is a terminal and $\nsymc{\sigma}$ is a (possibly empty) word
% of nonterminals.
A grammar in Greibach normal form is said to be
simple~\cite{DBLP:conf/focs/KorenjakH66} if, for every nonterminal $\nsymc{X}$
and every terminal $\tsymc{a}$, there is at most one production of the
form $\gprod{X}{\tsymc{a}\nsymc{\gamma}}$.

We define a notion of bisimulation for grammars in Greibach normal form via a
labelled transition system. The system comprises a set of states $\N^\ast$
corresponding to words of nonterminal symbols. For each production
$\gprod{X}{\tsymc{a}\nsymc{\gamma}}$ and each word of nonterminal symbols
$\nsymc{\delta}$, we have a labelled transition
$\gltsred{X\delta}{a}{\gamma\delta}$. We let $\gequiv$ denote the bisimulation
relation for grammars in Greibach normal form.
%
% Given a nonterminal word $\nsymc{\sigma}$ and a terminal word $\tsymc{\alpha}$, we define the image of $\tsymc{\alpha}$ on $\nsymc{\sigma}$, denoted by $\nsymc{\sigma}\cdot \tsymc{\alpha}$, recursively on $\tsymc{\alpha}$ as follows.
%
% \begin{itemize}
% 	\item For the empty word $\tsymc{\varepsilon}$, we have that $\nsymc{\sigma}\cdot \tsymc{\varepsilon} = \nsymc{\sigma}$.
% 	\item If $\tsymc{\alpha} = \tsymc{a\alpha'}$ is non-empty, then $\nsymc{\sigma}\cdot \tsymc{\alpha}$ is defined iff $\nsymc{\sigma}$ is of the form $\nsymc{X}{\nsymc{\sigma''}}$, there is a production $\gprod{X}{\tsymc{a}\nsymc{\sigma'}}$, and $\nsymc{\sigma'\sigma''}\cdot \tsymc{\alpha'}$ is defined; in which case, $\nsymc{\sigma}\cdot \tsymc{\alpha} = \nsymc{\sigma'\sigma''}\cdot \tsymc{\alpha'}$.
% \end{itemize}
%
% For simple grammars, whenever $\nsymc{\sigma}\cdot \tsymc{\alpha}$ is defined, it is uniquely defined. In such cases we say that $\tsymc{\alpha}$ is a trace of $\nsymc{\sigma}$. We define the language of traces of $\nsymc{\sigma}$ as
%
% $$\lcal(\nsymc{\sigma}) = \{\tsymc{\alpha} : \nsymc{\sigma}\cdot \tsymc{\alpha} \text{ is defined} \}.$$
%
% \todo{I think we would like to refer to grammars bisimulation instead of language equality.}
%
% \paragraph{Translation of types into grammars.}

Our next step is to explain how to convert a type into a simple grammar. We do
this in the two steps outlined below. For any type $\typec{T}$, let
$\subexpr(\typec{T})$ denote the set of subterms of $\typec{T}$.

\myparagraph{Step 1. Construct a grammar} Suppose we have a type
$\TT$ defined by means of a signature
$\Sigma=\{\iseqt{X_i}{T_i}\}_{i=1\ldots m}$. For every subterm $\typec U$
appearing in $\TT$ as well as in the equations in $\Sigma$, let
$\nsymc{X}_{\nsymc U}$ denote a fresh nonterminal symbol. Moreover, let
$\nsymc\bot$ denote a nonterminal symbol distinct from all
$\nsymc{X}_{\nsymc U}$. We define the grammar $(\T, \N, \nsymc{X}_{\nsymc T}, \R)$
where $\T$ is the language of the transition labels in \cref{fig:lts}, $\N$ is
the set
\begin{equation*}
  %   \T &= \{\tsymc{\keyword{unit}}, \ldomain, \tsymc{\lrange}, 
  %   \tsymc{?_d}, \tsymc{?_c}, \tsymc{!_d}, \tsymc{!_c}, 
  %   %\tsymc{\times_l}, \tsymc{\times_r}, 
  %   \tsymc{\forall^\tkind}, \tsymc{\exists^\tkind},\tsymc{\forall^\skind}, \tsymc{\exists^\skind}\}
  %   \cup
  %   \{\tsymc{n} : n \in \nbb\}
  %   \cup
  %   \bigcup\nolimits_{\ell\in\mathbb{L}}\{\tsymc{\records{}_\ell},
  %        \tsymc{\variants{}_\ell}, \tsymc{\&_\ell}, \tsymc{\oplus_\ell}\}
  % \\
  %   \N &=
  \{\nsymc{\bot}\} \cup
  \{\nsymc{X}_{\nsymc U} : \typec U \in \subexpr(\TT)\}\cup
  \bigcup\nolimits_{i=1\ldots m}\{\nsymc{X}_{\nsymc U} : \typec U \in \subexpr(\typec{T_i})\}
\end{equation*}
and the productions in $\R$ for nonterminal $\nsymc{X}_{\nsymc U}$ are defined
according to the syntax of $\typec U$ as follows

\smallskip\noindent
\begin{multicols}{2}
  \begin{center}
    \begin{tabular}[c]{cl}
      Type $\typec U$ & Productions for $\nsymc{X}_{\nsymc U}$
      \\
      \hline
      $\unit$ & $\gprod{\nsymc{X}_{\nsymc U}}{\tsymc{\unitl}}$
      \\
      $\function{V}{W}$ & 
      $\gprod{\nsymc{X}_{\nsymc U}}{\ldomain\nsymc{X}_{\nsymc{V}}}$,\quad
      $\gprod{\nsymc{X}_{\nsymc U}}{\lrange\nsymc{X}_{\nsymc{W}}}$
      \\
      $\typec{\varrecf{\ell}{U_\ell}{L}}$ &
      $\gprod{\nsymc{X}_{\nsymc U}}{\tsymc{\varrecs{}_k}\nsymc{X}_{\nsymc{U_k}}}$,
      for each $k\in L$
      \\
      $\typec{\quant^\vkind{V}}$ & $\gprod{\nsymc{X}_{\nsymc U}}{\tsymc{\quant^{\tsymc{\vkind}}}\nsymc{X}_{\nsymc{V}}}$
      \\
      $\Skip$ & $\gprod{\nsymc{X}_{\nsymc U}}{\nsymc{\Empty}}$
    \end{tabular}
  \end{center}
  \begin{center}
    \begin{tabular}[c]{cl}
      Type $\typec U$ & Productions for $\nsymc{X}_{\nsymc U}$
      \\
      \hline
      $\typec{\sharp{V}}$ &
      $\gprod{\nsymc{X}_{\nsymc U}}{\tsymc{\sharp_d}\nsymc{X}_{\nsymc{V}}\nsymc{\bot}}$,\quad
      $\gprod{\nsymc{X}_{\nsymc U}}{\tsymc{\sharp_c}}$
      \\
      $\typec{\choice\record{\ell}{U}{L}}$ &
      $\gprod{\nsymc{X}_{\nsymc U}}{\tsymc{\choicep{}_k}\nsymc{X}_{\nsymc{U_k}}}$,
      for each $k\in L$
      \\
      $\semit{V}{W}$ & $\gprod{\nsymc{X}_{\nsymc U}}{\nsymc{X}_{\nsymc{V}}\nsymc{X}_{\nsymc{W}}}$
      \\
      $\typec n$ & $\gprod{\nsymc{X}_{\nsymc U}}{\tsymc{n}}$
      \\
      $\typec{X_i}$ & $\gprod{\nsymc{X}_{\nsymc U}}{\nsymc{X}_{\nsymc{T_i}}}$
  \end{tabular}
\end{center}
\end{multicols}

Notice that  symbol $\nsymc{\bot}$ has no production.

\myparagraph{Step 2. Transform into a simple grammar}
In this step we convert the grammar constructed in the previous step into
Greibach normal form. We need to take care of the productions
$\gprod{\nsymc{X}}{\nsymc{\Empty}}$, $\gprod{\nsymc{X}}{\nsymc{Y}}$, and
$\gprod{\nsymc{X}}{\nsymc{Y}\nsymc{Z}}$ which are not in Greibach normal form.
First, suppose we have a production $\gprod{\nsymc{X}}{\nsymc{\Empty}}$; by
construction, this is the only such production for $\nsymc{X}$. We remove every
production of the form $\gprod{\nsymc{X}}{\nsymc{\Empty}}$ from our grammar and erase
each such $\nsymc{X}$ from the right-hand side of every production where it
appears. Next, suppose we have a production
$\gprod{\nsymc{X}}{\nsymc{Y}\nsymc{\gamma}}$ where $\nsymc{\gamma}$ is a
(possibly empty) word of nonterminal symbols.
% ; by construction, this is the only such production for $\nsymc{X}_\typec{U}$. 
We remove this production and, for each production $\gprod{\nsymc{Y}}{\sigma}$, include a production $\gprod{\nsymc{X}}{\sigma\nsymc{\gamma}}$. 
%By construction, either all productions $\gprod{\nsymc{X}_\typec{V}}{\sigma}$ have a leading terminal symbol in $\sigma$, in which case the resulting productions are in Greibach normal form; or there is only one production of the form $\gprod{\nsymc{X}_\typec{V}}{\nsymc{\sigma}}$ (and $\nsymc{\sigma}$ is a string of nonterminals), in which case we get only one production $\gprod{\nsymc{X}_\typec{U}}{\nsymc{\sigma\gamma}}$. 
We continue in this fashion until all productions are in Greibach normal form.%; there is no way to be stuck in a cycle since our system is contractive.
% \end{description}

\begin{restatable}{proposition}{grammarconstruction}
\label{prop:typetogrammar}
	For any type $\typec{T}$ described by a signature $\Sigma$, the construction outlined in \cref{sec:grammars} terminates yielding a simple grammar.
\end{restatable}

The above construction introduces a nonterminal symbol $\nsymc{\bot}$ without
productions. Intuitively, $\nsymc{\bot}$ is used to separate the two descendants
of a send/receive operation. A type $\semit {\OUTn{T}}U$ must have a data
transition $\tsymc{!_d}$ to $\typec T$ and a continuation transition
$\tsymc{!_c}$ to $\typec U$. It must have two different transitions, since we
want to distinguish $\semit {\OUTn{T}}U$ from $\OUTn{(\semit TU)}$. For example,
the type $\semit {\OUTn\Skip}{\OUTn\Skip}$ sends two (empty) channels in
sequence, whereas $\OUTn{(\semit\Skip{\OUTn\Skip})}$ sends a channel which in
turns sends an empty channel. Moreover, when transitioning to the data
$\typec T$ of a sequential composition $\semit{\OUTn T}U$, we want to make sure
that we follow the grammar corresponding only to $\typec T$. The following
example provides some more insight. Suppose we have types $\typec T$, $\typec U$
given by  equations
\begin{equation*}
  \iseqt T {\semit{\OUTn V}W}
  \qquad
  \iseqt U {\semit{\OUTn {(\semit VV)}}W}
  \qquad
  \iseqt V{\intchoice\records{\tkeyword{go}\colon \Skip}}
  \qquad
  \iseqt W{\intchoice\records{\tkeyword{go}\colon W}}
\end{equation*}
Notice that $\isnotequiv TU$, as the type being sent in $\typec T$ offers a choice only once, whereas the type being sent in $\typec U$ offers that choice twice. Following the construction above, we arrive at the grammar with productions
$\gltsred{X_\nsymc{T}}{\tsymc{!_d}}{X_\nsymc{V}\bot X_\nsymc{W}}$, $\gltsred{X_\nsymc{T}}{!_c}{X_\nsymc{W}}$,
$\gltsred{X_\nsymc{U}}{!_d}{X_\nsymc{V}X_\nsymc{V}\bot X_\nsymc{W}}$, $\gltsred{X_\nsymc{U}}{!_c}{X_\nsymc{W}}$, $\gltsred{X_\nsymc{V}}{\oplus\symkeyword{go}}{\varepsilon}$, $\gltsred{X_\nsymc{W}}{\oplus\symkeyword{go}}{X_\nsymc{W}}.$
Now we can check that $\nsymc{X}_\nsymc{T}\not\approx\nsymc{X}_\nsymc{U}$, as
$$\gltsred{X_\nsymc{T}}{!_d}{X_\nsymc{V}\bot X_\nsymc{W}}\gltsred{}{\oplus\symkeyword{go}}{\bot X_\nsymc{W}}\not\longrightarrow
\quad\text{but}\quad
\gltsred{X_\nsymc{U}}{!_d}{X_\nsymc{V}X_\nsymc{V}\bot X_\nsymc{W}}\gltsred{}{\oplus\symkeyword{go}}{X_\nsymc{V}\bot X_\nsymc{W}}\gltsred{}{\oplus\symkeyword{go}}{\bot X_\nsymc{W}}.$$
Suppose instead that we did not have the nonterminal $\nsymc\bot$. In this case
we would have productions \mbox{$\gltsred{X_\nsymc{T}}{!_d}{X_\nsymc{V}X_\nsymc{W}}$} and $\gltsred{X_\nsymc{U}}{!_d}{X_\nsymc{V}X_\nsymc{V}X_\nsymc{W}}$ instead. Because $\typec W$ is an infinitely repeating type, we would undesirably conclude that $\nsymc{X}_\nsymc{T}\gequiv\nsymc{X}_\nsymc{U}$; in particular, we would have the infinite sequences of transitions
$$\gltsred{X_\nsymc{T}}{!_d}{X_\nsymc{V}X_\nsymc{W}}\gltsred{}{\oplus\symkeyword{go}}{X_\nsymc{W}}\gltsred{}{\oplus\symkeyword{go}}{X_\nsymc{W}}\gltsred{}{\oplus\symkeyword{go}}{\cdots}
\quad\text{and}\quad
\gltsred{X_\nsymc{U}}{!_d}{X_\nsymc{V}X_\nsymc{V}X_\nsymc{W}}\gltsred{}{\oplus\symkeyword{go}}{X_\nsymc{V}X_\nsymc{W}}\gltsred{}{\oplus\symkeyword{go}}{X_\nsymc{W}}\gltsred{}{\oplus\symkeyword{go}}{\cdots}.$$

\begin{restatable}[Soundness and completeness for grammars]{theorem}{soundcompletegrammar}
\label{thm:soundnesscompletenessgrammars}
Let $\typec{T}$, $\typec{U}$ be types and $(\T_{\typec{T}},\N_{\typec{T}},\nsymc{X}_{\nsymc{T}},\R_{\typec{T}})$, $(\T_{\typec{U}},\N_{\typec{U}},\nsymc{X}_{\nsymc{U}},\R_{\typec{U}})$ the corresponding simple grammars obtained by the construction outlined in \cref{sec:grammars}. Then $\isequiv TU$ iff $\nsymc{X}_{\nsymc{T}}\gequiv\nsymc{X}_{\nsymc{U}}$.
\end{restatable}

\section{An algorithm to decide type equivalence}
\label{sec:algorithm}

We are now in a position to describe the algorithm to decide type equivalence.
Our algorithm builds on the construction outlined in \cref{sec:grammars}, as
well as on a procedure for deciding bisimulation of simple grammars. Almeida
\etal~\cite{DBLP:conf/tacas/AlmeidaMV20} describe one such algorithm, which
incidentally equips the FreeST programming language~\cite{freest}. See
\cref{sec:related-work} for alternative algorithms for checking the bisimilarity
of simple grammars.

\begin{description}
\item[Input:] Two types $\TT$, $\typec U$ built on a common signature
  $\Sigma$.
\item[Output:] `YES' if $\isequiv TU$, `NO' otherwise.
\item[Algorithm:]\
  \begin{enumerate}
  \item Construct the simple grammar
    $(\T_{\typec{T}},\N_{\typec{T}},\nsymc{X}_{\nsymc{T}},\R_{\typec{T}})$
    corresponding to $\TT$.
  \item Construct the simple grammar
    $(\T_{\typec{U}},\N_{\typec{U}},\nsymc{X}_{\nsymc{U}},\R_{\typec{U}})$
    corresponding to $\typec U$.
  \item Use a decision algorithm to decide whether
    $\nsymc{X}_{\nsymc{T}}\gequiv\nsymc{X}_{\nsymc{U}}$.
  \end{enumerate}
\end{description}

\begin{restatable}{theorem}{termination}
  \label{thm:termination}
  The type equivalence algorithm terminates.
  Its computational complexity is doubly exponential.
\end{restatable}

% \begin{proof}
% Steps 1 and 2 of the above algorithm terminate due to \cref{prop:typetogrammar}. The number of terminal and nonterminal symbols of the resulting grammars is bounded by the number of subexpressions in $\TT$, $\typec U$ and $\Sigma$, which in turn is linear in the input size. The procedure to convert the grammar into simple grammar essentially amounts to `shortcutting' paths, which is polynomial in the number of nonterminals. Step 3 has the worst complexity bound, since it is doubly-exponential on the size of the grammars obtained in steps 1 and 2, which in turn is linear on the input size. Therefore, the entire algorithm has doubly-exponential running time.
% \end{proof}

\begin{restatable}{theorem}{soundcompletealgorithm}
\label{thm:soundnesscompletenessalgorithm}
The type equivalence algorithm is sound and complete with respect to $\teq$.
\end{restatable}
% \begin{proof}
% Let $\typec{T}$, $\typec{U}$ be context-free types and $\nsymc{X}_{\typec{T}}$, $\nsymc{X}_{\typec{U}},\R_{\typec{U}}$ the corresponding starting nonterminals as described above. From \cref{thm:soundnesscompletenessgrammars} we know that $\isequiv TU$ iff $\nsymc{X}_{\typec{T}}\gequiv\nsymc{X}_{\typec{U}}$. From the soundness and completeness of the Burkart \etal algorithm, we know that step 3 of our type equivalence algorithm correctly whether $\nsymc{X}_{\typec{T}}\gequiv\nsymc{X}_{\typec{U}}$. Therefore, $\isequiv TU$ iff the type equivalence algorithm return `YES'.
% \end{proof}
\begin{restatable}{corollary}{decidability}
\label{thm:decidability}
The type equivalence problem is decidable.
\end{restatable}

% \begin{proof}
% Direct consequence of \cref{thm:termination,thm:soundnesscompletenessalgorithm}.
% \end{proof}

%%% Local Variables:
%%% mode: latex
%%% TeX-master: "main"
%%% End:
% !TeX root = main.tex

\section{Related work}
\label{sec:related-work}

Related work is varied; we focus on that related to non-regular session types,
polymorphism and bisimulation checking algorithms.

\myparagraph{Beyond regular session types}
Since its original proposal in the 90s~\cite{DBLP:conf/concur/Honda93,DBLP:conf/esop/HondaVK98,DBLP:conf/parle/TakeuchiHK94}, the theory of session types has evolved substantially. 
The interest in non-regular protocols was already apparent~\cite{DBLP:conf/europar/Puntigam99,DBLP:conf/europar/RavaraV97,DBLP:conf/soco/Sudholt05} when Thiemann and Vasconcelos proposed context-free session types as a way to specify non-regular communication protocols~\cite{DBLP:conf/icfp/ThiemannV16}.
% Recent works have sought to extend this theory to types beyond regular~\cite{DBLP:conf/esop/DasDMP21,DBLP:journals/corr/abs-2201-08275,DBLP:conf/icfp/ThiemannV16}, but in the meantime several proposal had explored the use of non-regular protocols~\cite{DBLP:conf/europar/Puntigam99,DBLP:conf/europar/RavaraV97,DBLP:conf/soco/Sudholt05}. 
% Of more interest for this paper is the proposal of context-free session types by Thiemann and Vasconcelos~\cite{DBLP:conf/icfp/ThiemannV16}, 
Context-free session types were integrated in the FreeST programming
language~\cite{DBLP:journals/corr/abs-1904-01284} as soon as an implementation
for a type equivalence algorithm was
developed~\cite{DBLP:conf/tacas/AlmeidaMV20}. More recently, the language was
extended to System F~\cite{DBLP:journals/corr/abs-2106-06658}; here we follow
the same strategy and promote context-free session types to higher-order types.
An alternative implementation of context-free session type equivalence was
proposed by Padovani~\cite{DBLP:journals/toplas/Padovani19} by resorting to
explicit code annotations, thus greatly simplifying the decision problem.
Despite the interest of types characterized by context-free languages, types
that live beyond regular are not limited to context-free session types; Gay \etal~\cite{DBLP:journals/corr/abs-2201-08275} analyse different shades of session
types that go beyond regular.

\myparagraph{Polymorphic session types}
There has been a myriad of attempts to integrate polymorphic types into session types: from bounded polymorphism~\cite{DBLP:journals/mscs/Gay08}, to parametric and bounded polymorphism without recursion~\cite{DBLP:journals/iandc/DardhaGS17} or with recursion but without polymorphism~\cite{Dardha2014}; Wadler proposed the inclusion of explicit polymorphism~\cite{DBLP:conf/icfp/Wadler12}, 
which was then considered with parametric polymorphism but without general recursion~\cite{DBLP:conf/esop/CairesPPT13}, and afterwards with recursion but without nested 
 types~\cite{griffith2016polarized}. Finally, Das \etal proposed parametric polymorphism with nested types~\cite{DBLP:conf/esop/DasDMP21,DBLP:journals/corr/abs-2103-15193}. We propose an extension of System F types with higher-order context-free session types, taking advantage of polymorphic (functional) types, which is more closely related to polymorphic (first-order) context-free session types~\cite{DBLP:journals/corr/abs-2106-06658}.

\myparagraph{Semantic \textit{vs} syntatic approaches to type equivalence definition}
The syntactic method is the most common approach to type
equivalence~\cite{DBLP:books/daglib/0005958}.
% Semantic approaches however fit
% naturally in subtyping
% contexts~\cite{DBLP:conf/ppdp/CastagnaF05,DBLP:conf/lics/FrischCB02,DBLP:conf/types/PetruccianiCAZ18},
% from which notions of equivalence can be defined.
The first paper on sessions and recursion uses implicit equirecursive types, so
that one may classify type equivalence as a semantic
notion~\cite{DBLP:conf/esop/HondaVK98}. An explicit notion of type equivalence
for session types (actually of subtyping) was proposed by Gay and Hole, making
use of a bisimulation~\cite{DBLP:journals/acta/GayH05}. The same paper also
presents a syntactic, rule based definition as a basis for an algorithm. A
similar construction was used in building notions of equivalence for more
complex session types~\cite{DBLP:conf/esop/DasDMP21,DBLP:conf/icfp/ThiemannV16}.
This paper introduces both a syntactic and a semantic approach for a fairly rich language
of types.

% Several works have been trying to unify the syntactic and semantic approaches to
% subtyping, by providing sound and complete algorithmic rules. Ligatti \etal
% derived algorithmic rules for a subtyping relation on iso-recursive, inductively
% defined, types~\cite{DBLP:journals/toplas/LigattiBN17}. Zeeshan \etal derived
% subtyping rules from a semantic characterization of subtyping, in a mixed
% inductive and coinductive setting but also for the individual fragments of
% inductively and coinductively defined
% types~\cite{DBLP:journals/corr/abs-2201-10998}, but do not handle polymorphism
% and use simple (regular) recursion.

\myparagraph{Algorithms to decide the bisimilarity of simple grammars}
To the best of our knowledge, the only running algorithm for checking the
bisimilarity of simple grammars is that of Almeida
\etal\cite{DBLP:conf/tacas/AlmeidaMV20}.
Quite close to simple grammars, but inspired by concurrent processes, one finds
the basic process algebra (BPA)~\cite{DBLP:conf/parle/BaetenBK87}. BPA processes
were proven equivalent to grammars in Greibach normal form by Baeten
\etal~\cite{DBLP:journals/jacm/BaetenBK93}, so that decidability results and
algorithms for BPA may be readily transposed to grammars in Greibach normal form
(and hence to simple grammars).
Baeten \etal presented a decidability result for normed
BPA~\cite{DBLP:journals/jacm/BaetenBK93}, which was then extended to the full
BPA language by Christensen \etal~\cite{DBLP:journals/iandc/ChristensenHS95}. An
improved (elementary) algorithm was proposed by Burkart
\etal~\cite{DBLP:conf/mfcs/BurkartCS95}, and the complexity of this algorithm
was much later shown to be doubly exponential by Jan{\v
  c}ar~\cite{DBLP:journals/corr/abs-1207-2479}. For BPA processes, the
bisimilarity problem is known to be
EXPTIME-hard~\cite{DBLP:journals/ipl/Kiefer13}; however, this does not exclude
the possibility of a polynomial time algorithm for our model, since simple
grammars are less expressive than grammars in Greibach normal form. For a
different special case of normed BPA processes, Hirshfeld \etal presented a
polynomial-time algorithm for deciding
bisimilarity~\cite{DBLP:journals/tcs/HirshfeldJM96}.

% the algorithm for bisimulation of basic process algebras due to Burkart \etal~\cite{DBLP:conf/mfcs/BurkartCS95}. Their algorithm is an improvement of the algorithm by Christensen \etal~\cite{DBLP:journals/iandc/ChristensenHS95}, and follows the same ideas. It starts by building a finite base (\ie a relation from which we can derive a notion of equivalence) and then refines this base (by removing non-bisimilar pairs) until it arrives at a bisimulation-complete base. Their algorithm is known to run in doubly-exponential time \cite{DBLP:journals/corr/abs-1207-2479}. It was originally stated for basic process algebras, which are equivalent to grammars in Greibach normal form~\cite{DBLP:journals/jacm/BaetenBK93}.

%%% Local Variables:
%%% mode: latex
%%% TeX-master: "main"
%%% End:
% !TeX root = main.tex
\section{Conclusion}
\label{sec:conclusion}

This paper promotes context-free session types to the higher-order setting:
messages can now convey channels. We propose an extension of System F with
higher-order context-free session types and present three approaches for
defining type equivalence: a syntactic, rule-based version, a semantic version
based on bisimulations and an algorithmic version by reduction to simple grammar
bisimilarity. We show that the three formulations coincide. Algorithms exist for
deciding the bisimilarity of simple
grammars~\cite{DBLP:conf/tacas/AlmeidaMV20,DBLP:conf/mfcs/BurkartCS95}, from which
an algorithm for deciding type equivalence can be effectively constructed.
% We also prove the decidability of the equivalence problem for
% higher-order context-free session types.

Session types (kind $\skind$) are sometimes seen as special cases of functional
types (kind $\tkind$), meaning that a session type can be used in any context where a
functional type is
expected~\cite{DBLP:journals/corr/abs-2106-06658,freest,DBLP:journals/corr/abs-1904-01284}.
Such a notion can be captured by a subkind preorder generated by the
$\skind <: \tkind$ inequality. Likewise, types such as arrows, records or
variants can be tagged as linear or unrestricted (two alternative
multiplicities), depending on their intended usage. We plan to investigate the
incorporation of subkinding and multiplicities in the present system.

The polymorphic types we manipulate are functional, that is, type
$\forallt{\vkind}{T}$ is of kind $\tkind$. As such, type $\htreet$ with
$\iseqt {\htreet} {\intchoice\records{\nodel\colon
    \semit{\htreet}{\semit{\forallt{\tkind}{\OUTn0}}{\htreet}}, \leafl\colon
    \Skip}}$, streaming a binary tree of heterogeneous values, is considered ill
formed. This is a crucial assumption to guarantee that our translation yields a
simple grammar, as opposed to a context-dependent grammar. We plan to analyse
the implications of polymorphism over session types on the decidability of type
equivalence. % and on algorithms to check type equivalence.

Another challenge for future work is the incorporation of higher-order kinds.
The present kinds, $\skind$ and $\tkind$, are kinds of proper types. One can
also consider kinds for type families. For example, type $\dualoft$ is a type
constructor that, when given a session type, yields the dual session type.

%%% Local Variables:
%%% mode: latex
%%% TeX-master: "main"
%%% End:

\paragraph{Acknowledgements}
Support for this research was provided by the Fundação para a Ciência e a
Tecnologia through project SafeSessions, ref.\ PTDC/CCI-COM/6453/2020, by the
LASIGE Research Unit, ref.\ UIDB/00408/2020 and ref.\ UIDP/00408/2020, and 
by the COST Action CA20111.

\bibliographystyle{eptcs}
\bibliography{references}

\begin{thebibliography}{10}
\providecommand{\bibitemdeclare}[2]{}
\providecommand{\surnamestart}{}
\providecommand{\surnameend}{}
\providecommand{\urlprefix}{Available at }
\providecommand{\url}[1]{\texttt{#1}}
\providecommand{\href}[2]{\texttt{#2}}
\providecommand{\urlalt}[2]{\href{#1}{#2}}
\providecommand{\doi}[1]{doi:\urlalt{http://dx.doi.org/#1}{#1}}
\providecommand{\eprint}[1]{arXiv:\urlalt{https://arxiv.org/abs/#1}{#1}}
\providecommand{\bibinfo}[2]{#2}

\bibitemdeclare{article}{DBLP:journals/corr/abs-2106-06658}
\bibitem{DBLP:journals/corr/abs-2106-06658}
\bibinfo{author}{Bernardo \surnamestart Almeida\surnameend},
  \bibinfo{author}{Andreia \surnamestart Mordido\surnameend},
  \bibinfo{author}{Peter \surnamestart Thiemann\surnameend} \&
  \bibinfo{author}{Vasco~T. \surnamestart Vasconcelos\surnameend}
  (\bibinfo{year}{2021}): \emph{\bibinfo{title}{Polymorphic Context-free
  Session Types}}.
\newblock {\sl \bibinfo{journal}{CoRR}} \bibinfo{volume}{abs/2106.06658},
  \doi{10.48550/arXiv.2106.06658}.

\bibitemdeclare{misc}{freest}
\bibitem{freest}
\bibinfo{author}{Bernardo \surnamestart Almeida\surnameend},
  \bibinfo{author}{Andreia \surnamestart Mordido\surnameend} \&
  \bibinfo{author}{Vasco~T. \surnamestart Vasconcelos\surnameend}
  (\bibinfo{year}{2019}): \emph{\bibinfo{title}{{FreeST}, a Programming
  Language with Context-free Session Types}}.
\newblock \bibinfo{howpublished}{\url{http://rss.di.fc.ul.pt/tools/freest/}}.

\bibitemdeclare{inproceedings}{DBLP:journals/corr/abs-1904-01284}
\bibitem{DBLP:journals/corr/abs-1904-01284}
\bibinfo{author}{Bernardo \surnamestart Almeida\surnameend},
  \bibinfo{author}{Andreia \surnamestart Mordido\surnameend} \&
  \bibinfo{author}{Vasco~T. \surnamestart Vasconcelos\surnameend}
  (\bibinfo{year}{2019}): \emph{\bibinfo{title}{FreeST: Context-free Session
  Types in a Functional Language}}.
\newblock In: {\sl \bibinfo{booktitle}{PLACES}}, {\sl
  \bibinfo{series}{{EPTCS}}} \bibinfo{volume}{291}, pp.
  \bibinfo{pages}{12--23}, \doi{10.4204/EPTCS.291.2}.

\bibitemdeclare{inproceedings}{DBLP:conf/tacas/AlmeidaMV20}
\bibitem{DBLP:conf/tacas/AlmeidaMV20}
\bibinfo{author}{Bernardo \surnamestart Almeida\surnameend},
  \bibinfo{author}{Andreia \surnamestart Mordido\surnameend} \&
  \bibinfo{author}{Vasco~T. \surnamestart Vasconcelos\surnameend}
  (\bibinfo{year}{2020}): \emph{\bibinfo{title}{Deciding the Bisimilarity of
  Context-Free Session Types}}.
\newblock In: {\sl \bibinfo{booktitle}{TACAS}}, {\sl \bibinfo{series}{LNCS}}
  \bibinfo{volume}{12079}, \bibinfo{publisher}{Springer}, pp.
  \bibinfo{pages}{39--56}, \doi{10.1007/978-3-030-45237-7\_3}.

\bibitemdeclare{inproceedings}{DBLP:conf/parle/BaetenBK87}
\bibitem{DBLP:conf/parle/BaetenBK87}
\bibinfo{author}{Jos C.~M. \surnamestart Baeten\surnameend},
  \bibinfo{author}{Jan~A. \surnamestart Bergstra\surnameend} \&
  \bibinfo{author}{Jan~Willem \surnamestart Klop\surnameend}
  (\bibinfo{year}{1987}): \emph{\bibinfo{title}{Decidability of Bisimulation
  Equivalence for Processes Generating Context-Free Languages}}.
\newblock In: {\sl \bibinfo{booktitle}{PARLE}}, {\sl \bibinfo{series}{LNCS}}
  \bibinfo{volume}{259}, \bibinfo{publisher}{Springer}, pp.
  \bibinfo{pages}{94--111}, \doi{10.1007/3-540-17945-3\_5}.

\bibitemdeclare{article}{DBLP:journals/jacm/BaetenBK93}
\bibitem{DBLP:journals/jacm/BaetenBK93}
\bibinfo{author}{Jos C.~M. \surnamestart Baeten\surnameend},
  \bibinfo{author}{Jan~A. \surnamestart Bergstra\surnameend} \&
  \bibinfo{author}{Jan~Willem \surnamestart Klop\surnameend}
  (\bibinfo{year}{1993}): \emph{\bibinfo{title}{Decidability of Bisimulation
  Equivalence for Processes Generating Context-Free Languages}}.
\newblock {\sl \bibinfo{journal}{J. {ACM}}}
  \bibinfo{volume}{40}(\bibinfo{number}{3}), pp. \bibinfo{pages}{653--682},
  \doi{10.1145/174130.174141}.

\bibitemdeclare{inproceedings}{DBLP:conf/mfcs/BurkartCS95}
\bibitem{DBLP:conf/mfcs/BurkartCS95}
\bibinfo{author}{Olaf \surnamestart Burkart\surnameend},
  \bibinfo{author}{Didier \surnamestart Caucal\surnameend} \&
  \bibinfo{author}{Bernhard \surnamestart Steffen\surnameend}
  (\bibinfo{year}{1995}): \emph{\bibinfo{title}{An Elementary Bisimulation
  Decision Procedure for Arbitrary Context-Free Processes}}.
\newblock In: {\sl \bibinfo{booktitle}{MFCS}}, {\sl \bibinfo{series}{LNCS}}
  \bibinfo{volume}{969}, \bibinfo{publisher}{Springer}, pp.
  \bibinfo{pages}{423--433}, \doi{10.1007/3-540-60246-1\_148}.

\bibitemdeclare{inproceedings}{DBLP:conf/esop/CairesPPT13}
\bibitem{DBLP:conf/esop/CairesPPT13}
\bibinfo{author}{Lu{\'{\i}}s \surnamestart Caires\surnameend},
  \bibinfo{author}{Jorge~A. \surnamestart P{\'{e}}rez\surnameend},
  \bibinfo{author}{Frank \surnamestart Pfenning\surnameend} \&
  \bibinfo{author}{Bernardo \surnamestart Toninho\surnameend}
  (\bibinfo{year}{2013}): \emph{\bibinfo{title}{Behavioral Polymorphism and
  Parametricity in Session-Based Communication}}.
\newblock In: {\sl \bibinfo{booktitle}{ESOP}}, {\sl \bibinfo{series}{LNCS}}
  \bibinfo{volume}{7792}, \bibinfo{publisher}{Springer}, pp.
  \bibinfo{pages}{330--349}, \doi{10.1007/978-3-642-37036-6\_19}.

\bibitemdeclare{article}{DBLP:journals/iandc/ChristensenHS95}
\bibitem{DBLP:journals/iandc/ChristensenHS95}
\bibinfo{author}{S{\o}ren \surnamestart Christensen\surnameend},
  \bibinfo{author}{Hans \surnamestart H{\"{u}}ttel\surnameend} \&
  \bibinfo{author}{Colin \surnamestart Stirling\surnameend}
  (\bibinfo{year}{1995}): \emph{\bibinfo{title}{Bisimulation Equivalence is
  Decidable for All Context-Free Processes}}.
\newblock {\sl \bibinfo{journal}{Inf. Comput.}}
  \bibinfo{volume}{121}(\bibinfo{number}{2}), pp. \bibinfo{pages}{143--148},
  \doi{10.1006/inco.1995.1129}.

\bibitemdeclare{article}{Dardha2014}
\bibitem{Dardha2014}
\bibinfo{author}{Ornela \surnamestart Dardha\surnameend}
  (\bibinfo{year}{2014}): \emph{\bibinfo{title}{Recursive Session Types
  Revisited}}.
\newblock {\sl \bibinfo{journal}{EPTCS}} \bibinfo{volume}{162}, p.
  \bibinfo{pages}{27–34}, \doi{10.4204/eptcs.162.4}.

\bibitemdeclare{article}{DBLP:journals/iandc/DardhaGS17}
\bibitem{DBLP:journals/iandc/DardhaGS17}
\bibinfo{author}{Ornela \surnamestart Dardha\surnameend},
  \bibinfo{author}{Elena \surnamestart Giachino\surnameend} \&
  \bibinfo{author}{Davide \surnamestart Sangiorgi\surnameend}
  (\bibinfo{year}{2017}): \emph{\bibinfo{title}{Session types revisited}}.
\newblock {\sl \bibinfo{journal}{Inf. Comput.}} \bibinfo{volume}{256}, pp.
  \bibinfo{pages}{253--286}, \doi{10.1016/j.ic.2017.06.002}.

\bibitemdeclare{inproceedings}{DBLP:conf/esop/DasDMP21}
\bibitem{DBLP:conf/esop/DasDMP21}
\bibinfo{author}{Ankush \surnamestart Das\surnameend}, \bibinfo{author}{Henry
  \surnamestart DeYoung\surnameend}, \bibinfo{author}{Andreia \surnamestart
  Mordido\surnameend} \& \bibinfo{author}{Frank \surnamestart
  Pfenning\surnameend} (\bibinfo{year}{2021}): \emph{\bibinfo{title}{Nested
  Session Types}}.
\newblock In: {\sl \bibinfo{booktitle}{{ESOP}}}, {\sl \bibinfo{series}{LNCS}}
  \bibinfo{volume}{12648}, \bibinfo{publisher}{Springer}, pp.
  \bibinfo{pages}{178--206}, \doi{10.1007/978-3-030-72019-3\_7}.

\bibitemdeclare{article}{DBLP:journals/corr/abs-2103-15193}
\bibitem{DBLP:journals/corr/abs-2103-15193}
\bibinfo{author}{Ankush \surnamestart Das\surnameend}, \bibinfo{author}{Henry
  \surnamestart DeYoung\surnameend}, \bibinfo{author}{Andreia \surnamestart
  Mordido\surnameend} \& \bibinfo{author}{Frank \surnamestart
  Pfenning\surnameend} (\bibinfo{year}{2021}): \emph{\bibinfo{title}{Subtyping
  on Nested Polymorphic Session Types}}.
\newblock {\sl \bibinfo{journal}{CoRR}} \bibinfo{volume}{abs/2103.15193},
  \doi{10.48550/arXiv.2103.15193}.

\bibitemdeclare{inproceedings}{debruijn:1972:lambda}
\bibitem{debruijn:1972:lambda}
\bibinfo{author}{Nicolaas~Govert \surnamestart De~Bruijn\surnameend}
  (\bibinfo{year}{1972}): \emph{\bibinfo{title}{Lambda calculus notation with
  nameless dummies, a tool for automatic formula manipulation, with application
  to the Church-Rosser theorem}}.
\newblock In: {\sl \bibinfo{booktitle}{Indagationes Mathematicae}},
  \bibinfo{volume}{75}, \bibinfo{organization}{Elsevier}, pp.
  \bibinfo{pages}{381--392}, \doi{10.1016/1385-7258(72)90034-0}.

\bibitemdeclare{article}{DBLP:journals/mscs/Gay08}
\bibitem{DBLP:journals/mscs/Gay08}
\bibinfo{author}{Simon~J. \surnamestart Gay\surnameend} (\bibinfo{year}{2008}):
  \emph{\bibinfo{title}{Bounded polymorphism in session types}}.
\newblock {\sl \bibinfo{journal}{MSCS}}
  \bibinfo{volume}{18}(\bibinfo{number}{5}), pp. \bibinfo{pages}{895--930},
  \doi{10.1017/S0960129508006944}.

\bibitemdeclare{article}{DBLP:journals/acta/GayH05}
\bibitem{DBLP:journals/acta/GayH05}
\bibinfo{author}{Simon~J. \surnamestart Gay\surnameend} \&
  \bibinfo{author}{Malcolm \surnamestart Hole\surnameend}
  (\bibinfo{year}{2005}): \emph{\bibinfo{title}{Subtyping for session types in
  the pi calculus}}.
\newblock {\sl \bibinfo{journal}{Acta Informatica}}
  \bibinfo{volume}{42}(\bibinfo{number}{2-3}), pp. \bibinfo{pages}{191--225},
  \doi{10.1007/s00236-005-0177-z}.

\bibitemdeclare{article}{DBLP:journals/corr/abs-2201-08275}
\bibitem{DBLP:journals/corr/abs-2201-08275}
\bibinfo{author}{Simon~J. \surnamestart Gay\surnameend}, \bibinfo{author}{Diogo
  \surnamestart Po{\c{c}}as\surnameend} \& \bibinfo{author}{Vasco~T.
  \surnamestart Vasconcelos\surnameend} (\bibinfo{year}{2022}):
  \emph{\bibinfo{title}{The Different Shades of Infinite Session Types}}.
\newblock {\sl \bibinfo{journal}{CoRR}} \bibinfo{volume}{abs/2201.08275},
  \doi{10.48550/arXiv.2201.08275}.

\bibitemdeclare{incollection}{girard:1971:extension}
\bibitem{girard:1971:extension}
\bibinfo{author}{Jean-Yves \surnamestart Girard\surnameend}
  (\bibinfo{year}{1971}): \emph{\bibinfo{title}{Une extension de
  L'interpretation de G{\"o}del a L'analyse, et son application a L'elimination
  des coupures dans L'analyse et la theorie des types}}.
\newblock In: {\sl \bibinfo{booktitle}{Studies in Logic and the Foundations of
  Mathematics}}, \bibinfo{volume}{63}, \bibinfo{publisher}{Elsevier}, pp.
  \bibinfo{pages}{63--92}, \doi{10.1016/S0049-237X(08)70843-7}.

\bibitemdeclare{article}{greibach:1965:normalform}
\bibitem{greibach:1965:normalform}
\bibinfo{author}{Sheila~A. \surnamestart Greibach\surnameend}
  (\bibinfo{year}{1965}): \emph{\bibinfo{title}{A New Normal-Form Theorem for
  Context-Free Phrase Structure Grammars}}.
\newblock {\sl \bibinfo{journal}{J. ACM}}
  \bibinfo{volume}{12}(\bibinfo{number}{1}), pp. \bibinfo{pages}{42–--52},
  \doi{10.1145/321250.321254}.

\bibitemdeclare{phdthesis}{griffith2016polarized}
\bibitem{griffith2016polarized}
\bibinfo{author}{Dennis~Edward \surnamestart Griffith\surnameend}
  (\bibinfo{year}{2016}): \emph{\bibinfo{title}{Polarized substructural session
  types}}.
\newblock Ph.D. thesis, \bibinfo{school}{University of Illinois at
  Urbana-Champaign}, \doi{10.2172/1562827}.

\bibitemdeclare{article}{DBLP:journals/tcs/HirshfeldJM96}
\bibitem{DBLP:journals/tcs/HirshfeldJM96}
\bibinfo{author}{Yoram \surnamestart Hirshfeld\surnameend},
  \bibinfo{author}{Mark \surnamestart Jerrum\surnameend} \&
  \bibinfo{author}{Faron \surnamestart Moller\surnameend}
  (\bibinfo{year}{1996}): \emph{\bibinfo{title}{A Polynomial Algorithm for
  Deciding Bisimilarity of Normed Context-Free Processes}}.
\newblock {\sl \bibinfo{journal}{Theor. Comput. Sci.}}
  \bibinfo{volume}{158}(\bibinfo{number}{1{\&}2}), pp.
  \bibinfo{pages}{143--159}, \doi{10.1016/0304-3975(95)00064-X}.

\bibitemdeclare{inproceedings}{DBLP:conf/concur/Honda93}
\bibitem{DBLP:conf/concur/Honda93}
\bibinfo{author}{Kohei \surnamestart Honda\surnameend} (\bibinfo{year}{1993}):
  \emph{\bibinfo{title}{Types for Dyadic Interaction}}.
\newblock In: {\sl \bibinfo{booktitle}{{CONCUR}}}, {\sl \bibinfo{series}{LNCS}}
  \bibinfo{volume}{715}, \bibinfo{publisher}{Springer}, pp.
  \bibinfo{pages}{509--523}, \doi{10.1007/3-540-57208-2\_35}.

\bibitemdeclare{inproceedings}{DBLP:conf/esop/HondaVK98}
\bibitem{DBLP:conf/esop/HondaVK98}
\bibinfo{author}{Kohei \surnamestart Honda\surnameend},
  \bibinfo{author}{Vasco~Thudichum \surnamestart Vasconcelos\surnameend} \&
  \bibinfo{author}{Makoto \surnamestart Kubo\surnameend}
  (\bibinfo{year}{1998}): \emph{\bibinfo{title}{Language Primitives and Type
  Discipline for Structured Communication-Based Programming}}.
\newblock In: {\sl \bibinfo{booktitle}{{ESOP}}}, {\sl \bibinfo{series}{LNCS}}
  \bibinfo{volume}{1381}, \bibinfo{publisher}{Springer}, pp.
  \bibinfo{pages}{122--138}, \doi{10.1007/BFb0053567}.

\bibitemdeclare{article}{DBLP:journals/corr/abs-1207-2479}
\bibitem{DBLP:journals/corr/abs-1207-2479}
\bibinfo{author}{Petr \surnamestart Jan{\v c}ar\surnameend}
  (\bibinfo{year}{2012}): \emph{\bibinfo{title}{Bisimilarity on Basic Process
  Algebra is in 2-ExpTime (an explicit proof)}}.
\newblock {\sl \bibinfo{journal}{Log. Methods Comput. Sci.}}
  \bibinfo{volume}{9}(\bibinfo{number}{1}), \doi{10.2168/LMCS-9(1:10)2013}.

\bibitemdeclare{article}{DBLP:journals/ipl/Kiefer13}
\bibitem{DBLP:journals/ipl/Kiefer13}
\bibinfo{author}{Stefan \surnamestart Kiefer\surnameend}
  (\bibinfo{year}{2013}): \emph{\bibinfo{title}{{BPA} bisimilarity is
  EXPTIME-hard}}.
\newblock {\sl \bibinfo{journal}{Inf. Process. Lett.}}
  \bibinfo{volume}{113}(\bibinfo{number}{4}), pp. \bibinfo{pages}{101--106},
  \doi{10.1016/j.ipl.2012.12.004}.

\bibitemdeclare{inproceedings}{DBLP:conf/focs/KorenjakH66}
\bibitem{DBLP:conf/focs/KorenjakH66}
\bibinfo{author}{A.~J. \surnamestart Korenjak\surnameend} \&
  \bibinfo{author}{John~E. \surnamestart Hopcroft\surnameend}
  (\bibinfo{year}{1966}): \emph{\bibinfo{title}{Simple Deterministic
  Languages}}.
\newblock In: {\sl \bibinfo{booktitle}{SWAT}}, \bibinfo{publisher}{{IEEE}
  Computer Society}, pp. \bibinfo{pages}{36--46}, \doi{10.1109/SWAT.1966.22}.

\bibitemdeclare{article}{DBLP:journals/toplas/Padovani19}
\bibitem{DBLP:journals/toplas/Padovani19}
\bibinfo{author}{Luca \surnamestart Padovani\surnameend}
  (\bibinfo{year}{2019}): \emph{\bibinfo{title}{Context-Free Session Type
  Inference}}.
\newblock {\sl \bibinfo{journal}{{ACM} Trans. Program. Lang. Syst.}}
  \bibinfo{volume}{41}(\bibinfo{number}{2}), pp. \bibinfo{pages}{9:1--9:37},
  \doi{10.1145/3229062}.

\bibitemdeclare{book}{DBLP:books/daglib/0005958}
\bibitem{DBLP:books/daglib/0005958}
\bibinfo{author}{Benjamin~C. \surnamestart Pierce\surnameend}
  (\bibinfo{year}{2002}): \emph{\bibinfo{title}{Types and programming
  languages}}.
\newblock \bibinfo{publisher}{{MIT} Press}.

\bibitemdeclare{inproceedings}{DBLP:conf/europar/Puntigam99}
\bibitem{DBLP:conf/europar/Puntigam99}
\bibinfo{author}{Franz \surnamestart Puntigam\surnameend}
  (\bibinfo{year}{1999}): \emph{\bibinfo{title}{Non-regular Process Types}}.
\newblock In: {\sl \bibinfo{booktitle}{Euro-Par}}, {\sl \bibinfo{series}{LNCS}}
  \bibinfo{volume}{1685}, \bibinfo{publisher}{Springer}, pp.
  \bibinfo{pages}{1334--1343}, \doi{10.1007/3-540-48311-X\_189}.

\bibitemdeclare{inproceedings}{DBLP:conf/europar/RavaraV97}
\bibitem{DBLP:conf/europar/RavaraV97}
\bibinfo{author}{Ant{\'{o}}nio \surnamestart Ravara\surnameend} \&
  \bibinfo{author}{Vasco~Thudichum \surnamestart Vasconcelos\surnameend}
  (\bibinfo{year}{1997}): \emph{\bibinfo{title}{Behavioural Types for a
  Calculus of Concurrent Objects}}.
\newblock In: {\sl \bibinfo{booktitle}{Euro-Par}}, {\sl \bibinfo{series}{LNCS}}
  \bibinfo{volume}{1300}, \bibinfo{publisher}{Springer}, pp.
  \bibinfo{pages}{554--561}, \doi{10.1007/BFb0002782}.

\bibitemdeclare{inproceedings}{DBLP:conf/programm/Reynolds74}
\bibitem{DBLP:conf/programm/Reynolds74}
\bibinfo{author}{John~C. \surnamestart Reynolds\surnameend}
  (\bibinfo{year}{1974}): \emph{\bibinfo{title}{Towards a theory of type
  structure}}.
\newblock In: {\sl \bibinfo{booktitle}{Programming Symposium}}, {\sl
  \bibinfo{series}{LNCS}}~\bibinfo{volume}{19}, \bibinfo{publisher}{Springer},
  pp. \bibinfo{pages}{408--423}, \doi{10.1007/3-540-06859-7\_148}.

\bibitemdeclare{book}{sangiorgi2014introduction}
\bibitem{sangiorgi2014introduction}
\bibinfo{author}{Davide \surnamestart Sangiorgi\surnameend}
  (\bibinfo{year}{2014}): \emph{\bibinfo{title}{An Introduction to Bisimulation
  and Coinduction}}.
\newblock \bibinfo{publisher}{Cambridge University Press}.

\bibitemdeclare{inproceedings}{DBLP:conf/soco/Sudholt05}
\bibitem{DBLP:conf/soco/Sudholt05}
\bibinfo{author}{Mario \surnamestart S{\"{u}}dholt\surnameend}
  (\bibinfo{year}{2005}): \emph{\bibinfo{title}{A Model of Components with
  Non-regular Protocols}}.
\newblock In: {\sl \bibinfo{booktitle}{SC}}, {\sl \bibinfo{series}{LNCS}}
  \bibinfo{volume}{3628}, \bibinfo{publisher}{Springer}, pp.
  \bibinfo{pages}{99--113}, \doi{10.1007/11550679\_8}.

\bibitemdeclare{inproceedings}{DBLP:conf/parle/TakeuchiHK94}
\bibitem{DBLP:conf/parle/TakeuchiHK94}
\bibinfo{author}{Kaku \surnamestart Takeuchi\surnameend},
  \bibinfo{author}{Kohei \surnamestart Honda\surnameend} \&
  \bibinfo{author}{Makoto \surnamestart Kubo\surnameend}
  (\bibinfo{year}{1994}): \emph{\bibinfo{title}{An Interaction-based Language
  and its Typing System}}.
\newblock In: {\sl \bibinfo{booktitle}{{PARLE}}}, {\sl \bibinfo{series}{LNCS}}
  \bibinfo{volume}{817}, \bibinfo{publisher}{Springer}, pp.
  \bibinfo{pages}{398--413}, \doi{10.1007/3-540-58184-7\_118}.

\bibitemdeclare{inproceedings}{DBLP:conf/icfp/ThiemannV16}
\bibitem{DBLP:conf/icfp/ThiemannV16}
\bibinfo{author}{Peter \surnamestart Thiemann\surnameend} \&
  \bibinfo{author}{Vasco~T. \surnamestart Vasconcelos\surnameend}
  (\bibinfo{year}{2016}): \emph{\bibinfo{title}{Context-free session types}}.
\newblock In: {\sl \bibinfo{booktitle}{ICFP}}, \bibinfo{publisher}{{ACM}}, pp.
  \bibinfo{pages}{462--475}, \doi{10.1145/2951913.2951926}.

\bibitemdeclare{inproceedings}{DBLP:conf/icfp/Wadler12}
\bibitem{DBLP:conf/icfp/Wadler12}
\bibinfo{author}{Philip \surnamestart Wadler\surnameend}
  (\bibinfo{year}{2012}): \emph{\bibinfo{title}{Propositions as sessions}}.
\newblock In: {\sl \bibinfo{booktitle}{{ICFP}}}, \bibinfo{publisher}{{ACM}},
  pp. \bibinfo{pages}{273--286}, \doi{10.1145/2364527.2364568}.

\end{thebibliography}

% \newpage
% \appendix
% %\setcounter{page}{1} % when latexing the appendices separately
% \input{appendix-proofs-typeequivalence}
% \input{appendix-proofs-grammar}

\end{document}